\documentclass[aps,superscriptaddress,prc,twocolumn,nofootinbib]{revtex4}

\usepackage{graphicx}
\usepackage{epstopdf}
\usepackage{subfigure}
\usepackage{epsfig}
\usepackage{amsmath,amssymb,amsfonts}
\usepackage{color}
\usepackage{url}
\usepackage[utf8]{inputenc}
\usepackage[bookmarks,
                bookmarksopen = true,
                bookmarksnumbered = true,
                linktocpage,
                colorlinks = true,
                linkcolor = blue,
                urlcolor  = blue,
                citecolor = blue,
                anchorcolor = green,
                hyperindex = true,
                hyperfigures]
                {hyperref}

\begin{document}

\title{QLBT: A linear Boltzmann transport model for heavy quarks in a quark-gluon plasma of quasi-particles}

\author{Feng-Lei Liu}
\affiliation{Institute of Particle Physics and Key Laboratory of Quark and Lepton Physics (MOE), Central China Normal University, Wuhan, 430079, China}

\author{Wen-Jing Xing}
\affiliation{Institute of Particle Physics and Key Laboratory of Quark and Lepton Physics (MOE), Central China Normal University, Wuhan, 430079, China}

\author{Xiang-Yu Wu}
\affiliation{Institute of Particle Physics and Key Laboratory of Quark and Lepton Physics (MOE), Central China Normal University, Wuhan, 430079, China}

\author{Guang-You Qin}
\email{guangyou.qin@mail.ccnu.edu.cn}
\affiliation{Institute of Particle Physics and Key Laboratory of Quark and Lepton Physics (MOE), Central China Normal University, Wuhan, 430079, China}

\author{Shanshan Cao}
\email{shanshan.cao@sdu.edu.cn}
\affiliation{Institute of Frontier and Interdisciplinary Science, Shandong University, Qingdao, Shandong 266237, China}

\author{Xin-Nian Wang}
\email{xnwang@lbl.gov}
\thanks{Current address: 3.}
\affiliation{Institute of Particle Physics and Key Laboratory of Quark and Lepton Physics (MOE), Central China Normal University, Wuhan, 430079, China}
\affiliation{Nuclear Science Division, Lawrence Berkeley National Laboratory, Berkeley, CA 94720, USA}

\date{\today}


\begin{abstract}

We develop a new heavy quark transport model, QLBT, to simulate the dynamical propagation of heavy quarks inside the quark-gluon plasma (QGP) created in relativistic heavy-ion collisions.
Our QLBT model is based on the linear Boltzmann transport (LBT) model with the ideal QGP replaced by a collection of quasi-particles to account for the non-perturbative interactions among quarks and gluons of the hot QGP.
The thermal masses of quasi-particles are fitted to the equation of state from lattice QCD simulations using the Bayesian statistical analysis method.
Combining QLBT with our advanced hybrid fragmentation-coalescence hadronization approach, we calculate the nuclear modification factor $R_\mathrm{AA}$ and the elliptic flow $v_2$ of $D$ mesons at the Relativistic Heavy-Ion Collider and the Large Hadron Collider.
By comparing our QLBT calculation to the experimental data on the $D$ meson $R_\mathrm{AA}$ and $v_2$, we extract the heavy quark transport parameter $\hat{q}$ and diffusion coefficient $D_\mathrm{s}$ in the temperature range of $1-4~T_\mathrm{c}$, and compare them with the lattice QCD results and other phenomenological studies.

\end{abstract}

\maketitle


\section{Introduction}
\label{sec:Introduction}

One of the main goals of relativistic heavy-ion collisions is to study the strong-interaction matter at extreme temperatures and densities, and to explore the properties of the quark-gluon plasma (QGP) in laboratories.
Experiments at Relativistic Heavy-Ion Collider (RHIC) and the Large Hadron Collider (LHC) have collected a tremendous amount of data with strong evidences for the formation of the color-deconfined QGP.
At low transverse momentum ($p_\mathrm{T}$), the observed hadron distributions exhibit large elliptic azimuthal anisotropies, which strongly depend on the centrality and collision geometry of the nucleus-nucleus collisions~\cite{Ollitrault:1992bk, Adler:2003kt, Adams:2003am, Aamodt:2010pa}.
Such azimuthal anisotropies can be successfully described by relativistic hydrodynamic model, implying a strongly-interacting QGP fluid  created in high-energy heavy-ion collisions at RHIC and the LHC.
Due to strong interaction among QGP constituents, the initial geometric anisotropies are converted into the anisotropic collective flow of the QGP and the final state momentum anisotropies of the produced hadrons~\cite{Gyulassy:1996br, Aguiar:2001ac, Broniowski:2007ft, Andrade:2008xh, Hirano:2009ah, Alver:2010gr, Petersen:2010cw, Qin:2010pf, Staig:2010pn, Teaney:2010vd, Schenke:2010rr, Ma:2010dv, Qiu:2011iv, Zhao:2019ehg}.
Currently, one important effort is to use the observed flow anisotropies to extract the specific shear viscosity of the QGP fluid via systematic comparisons with the relativistic hydrodynamics simulations~\cite{Heinz:2013th, Gale:2013da, Huovinen:2013wma, Bernhard:2019bmu, JETSCAPE:2020shq}.

At high $p_\mathrm{T}$, the observed hadron production in nucleus-nucleus collisions exhibits a strong suppression pattern compared to the expectation of independent nucleon-nucleon collisions~\cite{Khachatryan:2016odn, Acharya:2018qsh, Aad:2015wga, Burke:2013yra, Buzzatti:2011vt, Chien:2015vja, Andres:2016iys, Cao:2017hhk, Zigic:2018ovr}.
Such phenomenon is generally referred to as jet quenching~\cite{Wang:1991xy, Qin:2015srf, Blaizot:2015lma, Majumder:2010qh, Gyulassy:2003mc, Cao:2020wlm, Qin:2007rn}, which is mainly caused by energy loss of jet partons during their propagation through the QGP.
Jet partons may interact with the constituents of the hot medium and lose energy via elastic (collisional) and inelastic (radiative) interactions, before they fragment into color-neutral hadrons.
The interaction between jets and medium may also lead to the suppression of the production rates of fully reconstructed jets~\cite{Aad:2014bxa, Khachatryan:2016jfl, Qin:2010mn, Young:2011qx, Dai:2012am, Wang:2013cia, Blaizot:2013hx, Mehtar-Tani:2014yea, Cao:2017qpx, Kang:2017frl, He:2018xjv} and the modification of jet-related correlations~\cite{Aad:2010bu, Chatrchyan:2012gt, Qin:2009bk, Chen:2016vem, Chen:2016cof, Chen:2017zte, Luo:2018pto, Zhang:2018urd, Kang:2018wrs}.
It can also change the internal structures of full jets~\cite{Chatrchyan:2013kwa, Aad:2014wha, Chang:2016gjp, Casalderrey-Solana:2016jvj, Tachibana:2017syd, KunnawalkamElayavalli:2017hxo, Brewer:2017fqy, Chien:2016led, Milhano:2017nzm, Chang:2019sae} and induced medium response with interesting phenomenological consequences \cite{Qin:2009uh, Tachibana:2017syd, Yang:2021iib, Casalderrey-Solana:2020rsj, Chen:2020tbl, Yan:2017rku, Milhano:2017nzm, Gao:2016ldo}.
One important objective of the jet quenching study is to quantitatively determine various jet transport coefficients, such as $\hat{q}$ that characterizes the rate of transverse momentum broadening of jet partons inside the QGP~\cite{Burke:2013yra,Cao:2021keo}.

Heavy (charm and bottom) quarks are unique probes of QGP in relativistic heavy-ion collisions.
Due to their large masses ($m_Q \gg T_\mathrm{QGP}$), heavy quarks are produced at the early stage of nuclear collisions and thus are able to probe the entire history of the fireball evolution.
In addition, the production rate of heavy quarks can be computed via perturbative QCD techniques ($m_Q \gg \Lambda_\mathrm{QCD}$).
One of the compelling features of heavy quarks is the wide range of physics in the comprehensive coverage in $p_\mathrm{T}$.
At low $p_\mathrm{T}$, heavy quarks can be utilized to study the diffusion and thermalization processes inside the QGP.
At intermediate $p_\mathrm{T}$, heavy flavor hadrons are excellent tools to investigate the hadronization mechanisms of colored partons into color-neutral hadrons.
At high $p_\mathrm{T}$ where perturbative interactions dominate, heavy quarks can be used to investigate the flavor and mass dependences of parton energy loss and jet quenching.
Tremendous efforts have been devoted to the study of heavy quark dynamics in relativistic heavy-ion collisions~\cite{Dong:2019byy, Rapp:2018qla, Cao:2018ews, Uphoff:2011ad, He:2011qa, Young:2011ug, Alberico:2011zy, Nahrgang:2013saa, Cao:2013ita, Djordjevic:2013xoa, Cao:2015hia, Das:2015ana, Song:2015ykw, Cao:2016gvr, Kang:2016ofv, Prado:2016szr, Cao:2017crw, Liu:2017qah, Li:2018izm, Ke:2018tsh, Katz:2019fkc, Xing:2019xae, Li:2020kax}.

To describe heavy quark diffusion and energy loss in QGP, there are two common approaches: Langevin and Boltzmann transport models.
As for heavy quark hadronization, the combination of coalescence model and fragmentation mechanics is usually used; the former dominates in the lower and intermediate $p_\mathrm{T}$ regions while the later controls the high $p_\mathrm{T}$ region.
One of the most important tasks of heavy quark study is to obtain a simultaneous description of the experimental data on $R_\mathrm{AA}$ and $v_2$, and to quantitatively determine the spatial diffusion coefficient $D_\mathrm{s}(T)$ of heavy quarks. This is also the main purpose of our present work.

In this work, we improve the linear Boltzmann transport (LBT) model for heavy quark evolution in the QGP by modeling the QGP as a collection of thermalized quasi-particles.
In the LBT model, both elastic and inelastic interactions are included consistently for heavy quarks propagating through the QGP \cite{Wang:2013cia, He:2015pra, Cao:2016gvr, Cao:2017hhk, Xing:2019xae}.
In order to include the non-perturbative dynamics of the interaction among constituent quarks and gluons in the QGP, the properties (such as the thermal masses) of the quasi-particles are fitted to the realistic equation of state (EoS), i.e., the pressure, entropy, energy density as a function of temperature, as determined from lattice QCD calculations.
We call our new model the quasi-particle LBT (QLBT) model.
In the numerical calibration, we use the Bayesian statistical analysis to determine the values of the parameters for the quasi-particle masses (through the strong coupling $g$) by comparing the EoS from our thermal dynamical calculation to the lattice QCD results.
Two different sets of lattice QCD data, Wuppertal-Budapest (WB) and Hot QCD (HQ), are used and compared in our study.
After the parameters in the quasi-particle model are fixed, we then perform a systematic calculation of the $D$ meson $R_\mathrm{AA}$ and $v_2$ at RHIC and the LHC using our new QLBT model combined with our advanced hybrid fragmentation-coalescence hadronization approach.
By comparing our QLBT calculation to the corresponding data from CMS and STAR Collaborations, we determine the parameters in the interaction strength (strong coupling) between heavy quarks and the QGP medium using the Bayesian analysis method.
Using the  extracted parameters in our QLBT model, we further compute heavy quark transport parameter $\hat{q}$ and the spatial diffusion coefficients $D_\mathrm{s}$ of heavy quarks in the temperature range of $1-4~T_\mathrm{c}$.
Our numerical values of heavy quark diffusion coefficient as well as its temperature dependence are shown to be consistent with lattice QCD calculations and other phenomenological studies.

The remainder of this paper will be organized as follows. In Sec.~\ref{sec:QPM}, we will present our quasi-particle model of the thermal QGP. The comparison to lattice QCD equation of state and the extraction of our quasi-particle model parameters will be discussed. In Sec.~\ref{sec:QLBT}, we will review the (Q)LBT model which describes the evolution of heavy quarks inside QGP via elastic and inelastic interactions. In Sec.~\ref{sec:results}, we will present our numerical results on the $D$ meson $R_\mathrm{AA}$ and $v_2$ and extract the heavy quark transport coefficients with their temperature and momentum dependences. Our summary will be presented in Sec.~\ref{sec:summary}.

\section{The quasi-particle model of QGP}
\label{sec:QPM}

In the quasi-particle model, the system of interacting quarks and gluons in a thermalized QGP is effectively represented by an ideal gas of non-interacting massive quarks and quarks (called quasi-particles) \cite{Gorenstein:1995vm,Levai:1997yx,Bozek:1998dj,Bluhm:2004xn,Plumari:2011mk, Cassing:2008nn, Cassing:2008sv, Gossiaux:2009mk, Cassing:2009vt, Bratkovskaya:2011wp, Berrehrah:2014kba, Das:2015ana, Berrehrah:2015ywa, Song:2015ykw, Scardina:2017ipo}.
The masses of the quasi-particles contain all the non-perturbative dynamics of the interaction and can be viewed as the interaction energy among quarks and gluons in the QGP.
To describe the main thermodynamic features, such as the energy density, pressure and entropy density as a function of temperature, obtained from lattice QCD simulations, the thermal masses of quasi-particles have to be temperature dependent.
Usually a temperature-dependent bag pressure is also introduced to account for other non-perturbative effects and to meet the self-consistency of thermodynamic properties.

In this work, we utilize the following forms (motivated by perturative QCD calculation) for the temperature-dependent effective masses of quarks and gluons,
\begin{align}
  \label{eq:boltzmann1}
  & m_g^2(T) =\frac{1}{6} \left(N_c+\frac{1}{2} N_f \right)  g^2(T) T^2, \\
  & m_{q}^2(T) =\frac{N_c^2-1}{8N_c}g^2(T) T^2,
\end{align}
where $N_f$ is the number of flavors, $N_c$ is the number of colors, $m_g$ is the mass for gluons and $m_{q}$ is the mass of light quarks.

In the above expressions, the coupling factor $g(T)$ is temperature dependent, whose exact form will be fitted to the lattice QCD data. Again motivated by the perturbative QCD calculation, we use the following parametric form to model the temperature dependence of the coupling $g(T)$:
\begin{align}
g^2(T)=\frac{48 \pi^2}{(11 N_c-2 N_f) \ln \left[\frac{\left(a T/T_\mathrm{c}+b\right)^2 }{1+c e^{-d (T/T_\mathrm{c})^2 }}\right]}
\end{align}
where $a$, $b$, $c$ and $d$ are parameters to be determined by the equation of state from lattice QCD simulations.

\begin{figure*}[tbh]
\includegraphics[width=0.485\linewidth]{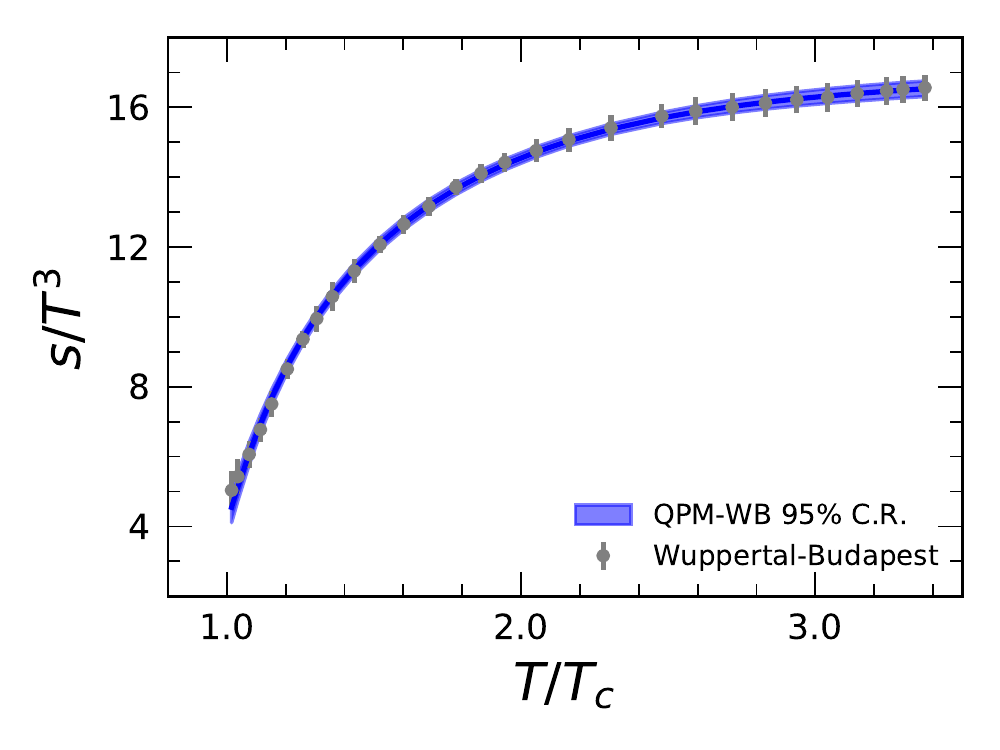}
\includegraphics[width=0.485\linewidth]{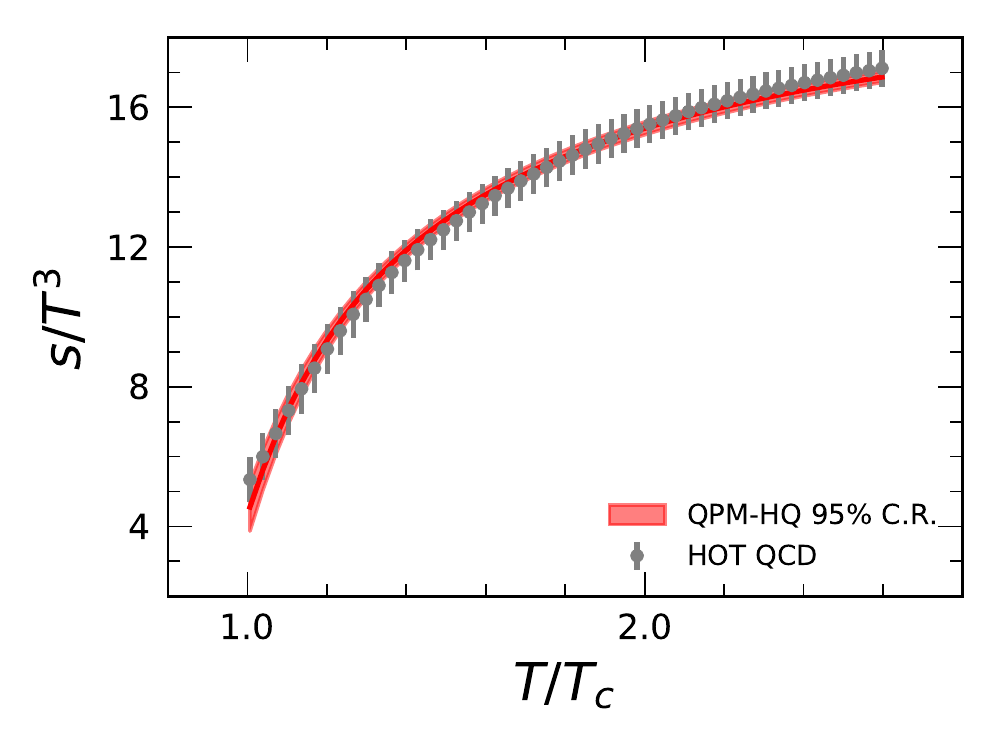}
\caption{The entropy density $s(T)$ from the quasi-particle model (QPM) is fitted to the lattice QCD data from Wuppertal-Budapest \cite{Borsanyi:2013bia} and the Hot QCD \cite{Bazavov:2014pvz} Collaborations. The bands show the $95\%$ confidence intervals.}
	\label{entropy-density}
\end{figure*}

With the ansatzs of $g(T)$ and the quasi-particle masses $m(T)$,  different thermodynamical quantities can be then calculated straightforwardly. First, the pressure can be calculated from the contributions of different constituents:
\begin{align}
  \label{eq:boltzmann1}
  P(T) &=\sum_{i} d_i \int \frac{d^3 p}{(2 \pi)^3}
  \frac{p^2}{3 E_i(p, T)} f_i(p, T)-B(T),
\end{align}
where the sum runs over all parton species, $f_i(p, T)$ denote Bose or Fermi distribution, $E_i(p, T) = \sqrt{p^2 + m_i^2(T)}$, $d_i$ is the spin and color degeneracy -- $2N_c$ for quarks (anti-quarks) and $2(N_c^2 - 1)$ for gluons, and $B(T)$ is temperature-dependent bag constant.
Following the strategy in Ref.~\cite{Plumari:2011mk}, the energy density of the system can be similarly obtained as follows:
\begin{align}
  \epsilon(T) &=\sum_{i} d_i \int \frac{d^3 p}{(2\pi)^3}E_i(p, T) f_i(p, T)+B(T).
\end{align}
As for the entropy density, the bag constant $B(T)$ in the pressure and energy density cancels:
\begin{align}
s(T) = \frac{\epsilon(T) + P(T)}{T}.
\end{align}
Therefore, the entropy density still preserves the ideal gas form. To minimize the uncertainty in the fitting procedure, we fix the parameters in $g(T)$ by fitting the lattice QCD data on the entropy density $s(T)$.

In order to extract the parameters in $g(T)$ from the lattice QCD data, we employ the Bayesian statistical analysis method, which can be simply summarized as:
\begin{align}
P(\theta | {\rm data}) = \frac{P(\theta) P({\rm data} | \theta)}{P({\rm data})}.
\end{align}
In the above equation, $P(\theta | {\rm data})$ is the posterior distribution of the parameter set $\theta =[a,b,c,d]$ given the experimental data, $P(\theta)$ is the prior distribution of the parameter set $\theta$, $P({\rm data}) = \int d\theta P(\theta) P({\rm data}|\theta)$ is the experimental evidence, and $P({\rm data} | \theta)$ is the Gaussian likelihood between experimental data and the output for any given set of parameters $\theta$:
\begin{align}
P({\rm data} | \theta) = \prod_i \frac{1}{\sqrt{2 \pi} \sigma_i} e^{- \frac{(y_i - y_\mathrm{exp})^2}{2 \sigma_i^2}},
\end{align}
where $y_i$ denotes model calculation results, $y_{\rm exp}$ denote the experiment or lattice QCD data, $\sigma_i$ denote the standard errors at each data point.
Following Ref.~\cite{He:2018gks}, we employ the PyMC library~\cite{pymc_bib} to perform the Markov Chain Monte Carlo (MCMC) estimation of the parameters with Metropolis-Hastings random walk in the parameter space.
During the $g(T)$ fitting process, the prior distributions of model parameters $(a,b,c,d)$ are taken as the uniform distribution within given ranges as shown in Table~\ref{tab:1}. The estimated values are then fed as the initial guess of these parameters in MCMC to sample $20000$ sets of parameters.

\begin{table}[htb]
\centering
\vspace{-5pt}
\begin{tabular}{c|c|c}
 \hline
 Lattice Data & Parameters & Prior Range  \\
 \hline
    & a & [0.1, 10]    \\
  WB   & b & [0, 1]  \\
      & c & [0, 10]   \\
      & d & [0, 2]   \\ \hline
   & a & [0, 10]   \\
  HQ   & b & [-1, 1]  \\
  	  & c & [10, 80]  \\
  	  & d & [0, 10]   \\
  \hline
\end{tabular}
	\caption{The ranges of model parameters $(a,b,c,d)$ used in the prior distributions.}
\label{tab:1}
\end{table}

In Fig.~\ref{entropy-density}, we present our calibration of the entropy density $s(T)$ as a function of temperature obtained from the quasi-particle model against the lattice QCD data from both the Wuppertal-Budapest (WB)~\cite{Borsanyi:2013bia} and the Hot QCD (HQ)~\cite{Bazavov:2014pvz} Collaborations. Note that for the two sets of lattice QCD data, the values of $T_\mathrm{c}$ are a little different: $T_\mathrm{c} = 150$~MeV for WB and $T_\mathrm{c} = 154$~MeV for HQ. To compare to different lattice results, we also use different $T_\mathrm{c}$ values in our quasi-particle model. In the figure, the bands represent the $95\%$ confidence region (C.R.) for the calibrated parameter values.

\begin{figure*}[tbh]
\includegraphics[width=0.485\linewidth]{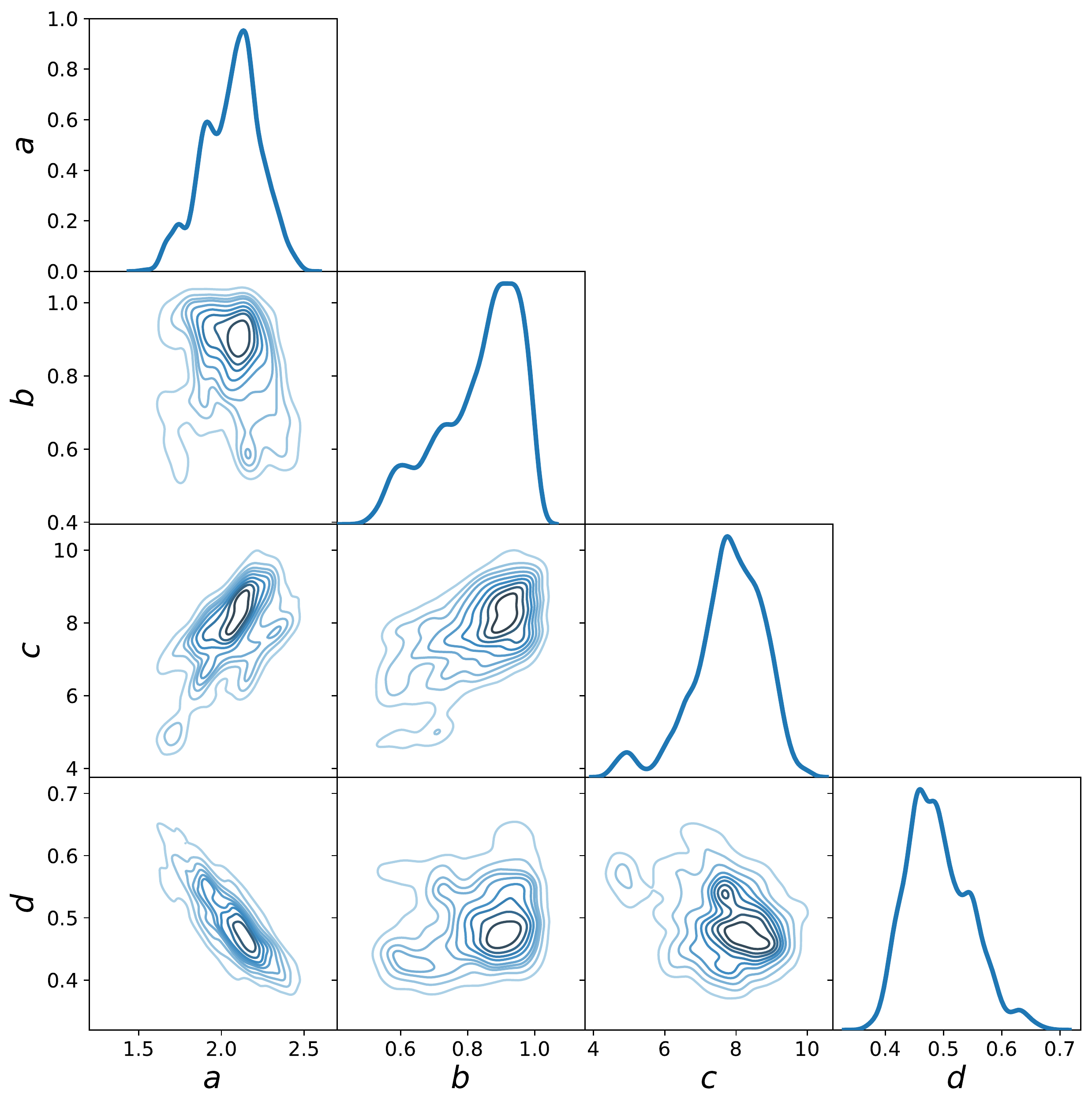}
\includegraphics[width=0.485\linewidth]{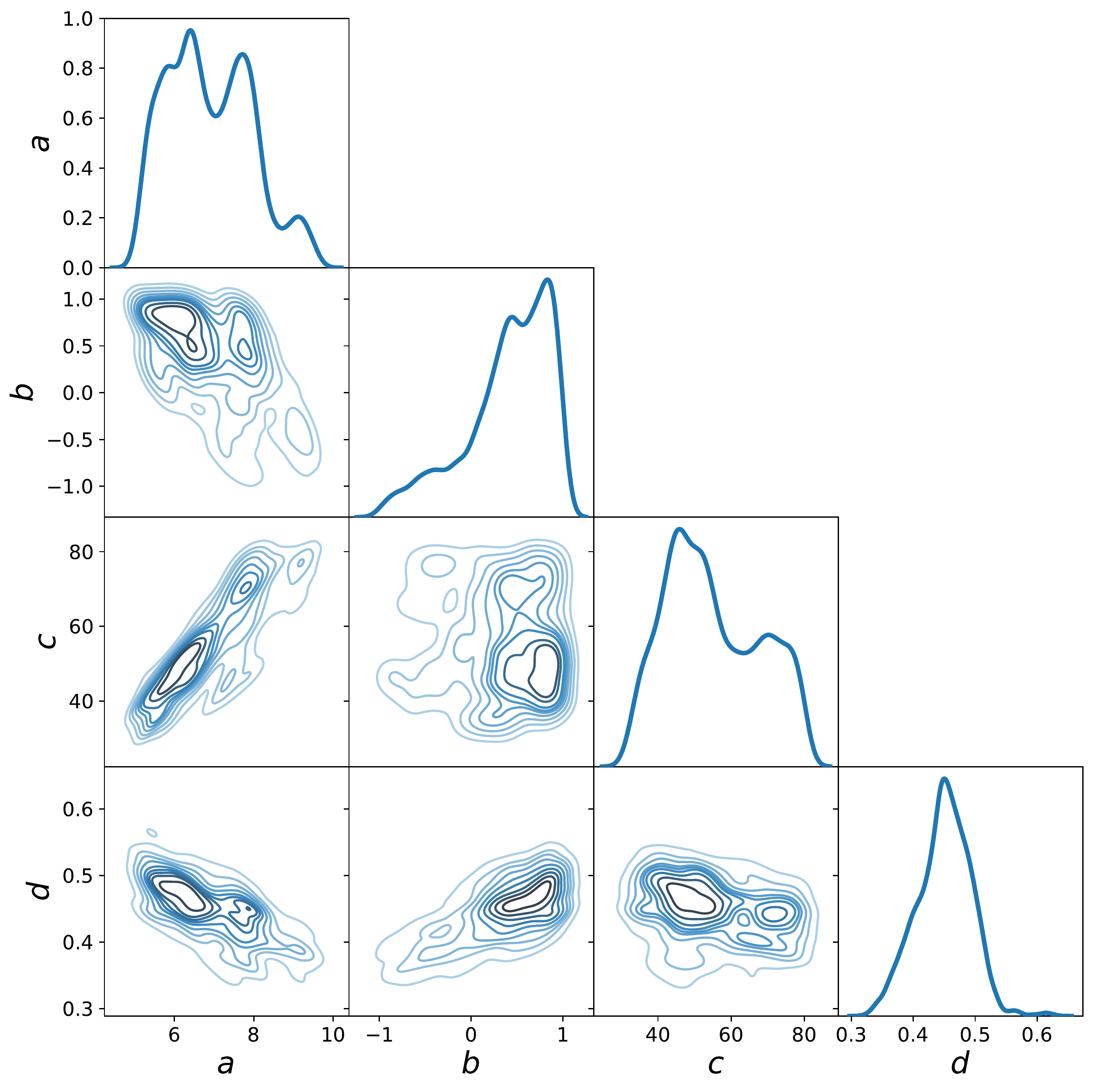}
\caption{The posterior probability distributions of the parameters $a$, $b$, $c$ and $d$ in the temperature-dependent coupling $g(T)$ by fitting to the lattice equation of state from Wuppertal-Budapest (Left) and Hot QCD (Right) collaobrations.}
	\label{abcd-values}
\vspace{12pt}
\includegraphics[width=0.485\linewidth]{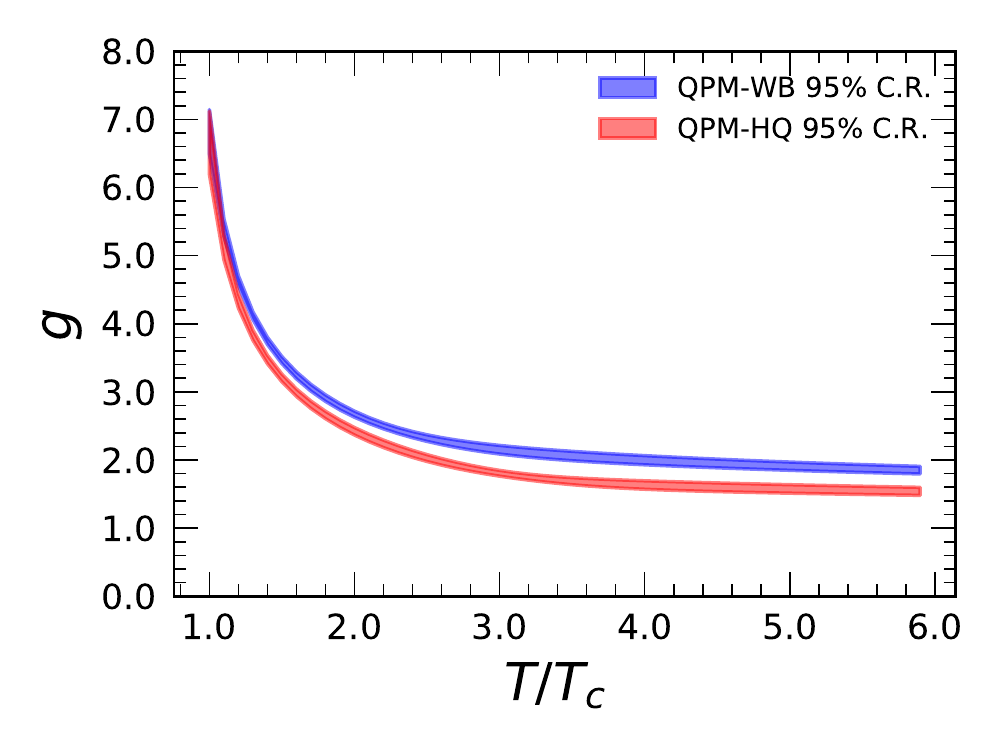}
\includegraphics[width=0.485\linewidth]{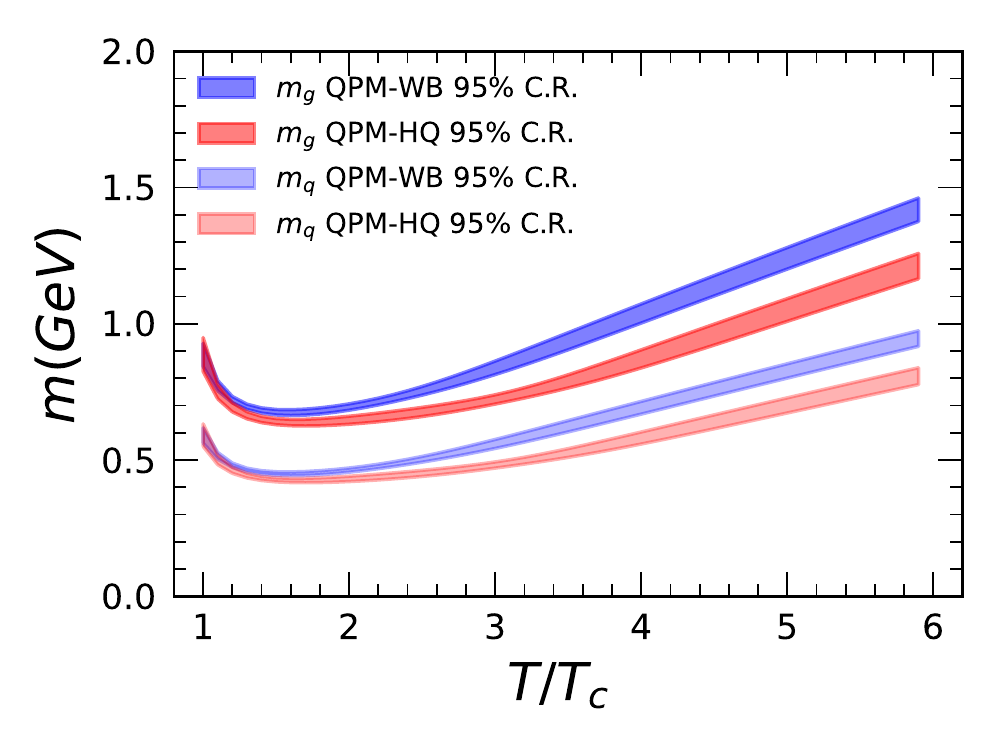}
	\caption{The strong coupling $g(T)$ (Left) and quasi-particle masses $m(T)$ (Right) using the fitted parameter values from the Bayesian analysis.}
	\label{gT-mT}
\end{figure*}

By comparing to the entropy density $s(T)$ as a function of temperature, we may extract the values of the parameters $a$, $b$, $c$ and $d$ in the temperature-dependent coupling $g(T)$. In Fig.~\ref{abcd-values}, we show the posterior distribution for these four parameters as well as the correlations between them. The left plot shows the calibration results using WB data, and the right plot for using HQ data.
The mean values and standard deviations from our calibration are summarized in Tab.~\ref{tab:2}.

\begin{table}[htb]
\label{tab:2}
\centering
\vspace{-5pt}
\begin{tabular}{c|c|c|c}
 \hline
 Lattice Data & Parameters & Mean Values & Standard Errors  \\
 \hline
    & a & 2.063 & 0.1705    \\
  WB   & b & 0.836 & 0.1186   \\
      & c & 7.792 & 0.9857   \\
      & d & 0.492 & 0.054   \\ \hline
   & a & 6.8899 & 1.055   \\
  HQ   & b & 0.398 & 0.462  \\
  	  & c & 54.9825 & 12.76  \\
  	  & d & 0.449 & 0.0432   \\
  \hline
\end{tabular}
	\caption{The mean and standard deviations of parameters $(a,b,c,d)$ in the coupling $g(T)$ extracted from Wuppertal-Budapest and Hot QCD lattice QCD data.}
\end{table}

Shown in Fig.~\ref{gT-mT} are the strong coupling $g(T)$ and quasi-particle masses $m(T)$ as functions of the medium temperature $T/T_\mathrm{c}$, in which the parameters $a$, $b$, $c$ and $d$ obtained from the above Bayesian analysis are applied.
The $95\%$ confidence interval bands are also shown.
One can observe from the left panel that the fitted coupling $g(T)$ shows a decreasing pattern as the temperature $T$ increases. When approaching $T_\mathrm{c}$, the coupling $g(T)$ can become very large.
In the fitting process, we disregard all the parameter sets which give divergent or negative values of $g(T)$.
In the right panel, one can see that as $T$ decreases, the quasi-particle masses first decrease and then increase; the transition is around $1.4T_\mathrm{c}$.
It is interesting to notice that the qualitative feature of strong coupling $g(T)$ and quasi-particle masses $m(T)$ are quite similar between the extractions from the two sets of lattice QCD data. However, quantitative difference still exists in $g(T)$ and $m(T)$ due to the different $T_\mathrm{c}$ values and temperature dependences of the entropy density $s(T)$ between WB and HQ data (as shown in Fig.~\ref{entropy-density}).

\section{(Q)LBT model for heavy quarks}
\label{sec:QLBT}

In the (Q)LBT model \cite{Wang:2013cia, He:2015pra, Cao:2016gvr, Cao:2017hhk, Xing:2019xae}, the evolution of the phase space distribution of a given parton (denoted as ``1" below) is described using the Boltzmann equation as follows:
\begin{align}
  \label{eq:boltzmann1}
  p_1\cdot\partial f_1(x_1,p_1)=E_1 (C_\mathrm{el} + C_\mathrm{inel}),
\end{align}
where $C_\mathrm{el}$ and $C_\mathrm{inel}$ are collision integrals arising from elastic and inelastic processes experienced by the propagating parton ``1".

For the elastic process ($1+2 \rightarrow 3+4$), the scattering rate is given by:
\begin{align}
\label{eq:gamma0}
\Gamma_{12 \to 34} (\vec{p}_1)
& = \frac{\gamma_2}{2E_1} \int \frac{d^3 p_2}{(2\pi)^3 2E_2} \int \frac{d^3 p_3}{(2\pi)^3 2E_3}\int \frac{d^3 p_4}{(2\pi)^3 2E_4}
\nonumber\\
& \times f_2(\vec{p}_2)[1\pm f_3(\vec{p}_3)] [1\pm f_4(\vec{p}_4)] S_2(s,t,u)
\nonumber\\
& \times (2\pi)^4 \delta^{(4)} (p_1 + p_2 - p_3 -p_4)|M_{12 \rightarrow 34}|^2,
\end{align}
where $\gamma_2$ is the spin-color degeneracy factor of parton ``2",  and is equal to $ 2 N_c$ for quarks (anti-quarks) and  $ 2 (N_c^2-1)$ for gluons.
In this work, we still keep the factor $S_2(s,t,u)=\theta(s\geq 2 \mu_D^2)\theta(t \leq-\mu_D^2)\theta(u \leq-\mu_D^2)$, with $\mu_D^2(T) = 2m_g^2(T)$, as in the case of massless thermal medium partons, where $S_2$ is imposed to avoid possible divergence at small angle $u, t \rightarrow 0$~\cite{Auvinen:2009qm, He:2015pra}. The leading order pQCD matrix elements are taken for $|M_{12 \rightarrow 34}|^2$ in elastic scattering process~\cite{Combridge:1978kx}. Note that the quasi-particle masses of the thermal partons have been included in evaluating the Mandelstam variables $s$, $t$ and $u$ that enter the matrix elements $|M_{12 \rightarrow 34}|^2$ although these elements were derived for scatterings between heavy quarks and massless partons~\cite{Combridge:1978kx}. The main effects of the quasi-particle masses in this work is on the density distribution of thermal partons $f_2(\vec{p}_2)$, which is significantly reduced compared to the density of massless partons. The sign $\pm$ represents the Pauli block and Bose enhancement effects.
The factor $(1-f_3)$ for heavy quarks is neglected considering that the temperature of QGP is much smaller than heavy quark masses, leading to very low heavy quark density inside the QGP.

Using the $\delta$-function and the azimuthal angular symmetry with respect to the $\vec{p}_1$ direction, Eq.~(\ref{eq:gamma0}) can be reduced to:
\begin{align}
\Gamma_{12\to 34} &=\frac{\gamma_2}{16 E_1 (2 \pi)^4}\int dE_2 d\theta_2 d\theta_4 d\phi_4 \nonumber\\
 &\times f_2(E_2,T)(1\mp f_4(E_4,T))S_2(s,t,u) \big| \mathcal{M_\mathrm{12\rightarrow34}} \big|^2
 \nonumber\\
 &\times \frac{p_2 p_4 \sin \theta_2 \sin \theta_4}{E_1+E_2-p_1 \cos \phi_4 \frac{E_4}{p_4}-p_2 \cos\theta_{24} \frac{E_4}{p_4}},
 \label{eq:gamma1}
\end{align}
where
\begin{align}
& \cos\theta_{24} = \sin \theta_2 \sin \theta_4 \cos \phi_4+ \cos \theta_2 \cos \theta_4,
\\
& E_4  = \frac{(E_1 + E_2)B \pm \sqrt{A^2 (m_4^2 A^2 + B^2 - m_4^2(E_1 + E_2)^2)}}{(E_1 + E_2)^2 - A^2}
\end{align}
with $A = |\vec{p}_1|\cos \theta_4 + |\vec{p}_2|\cos \theta_{24}$, and $B = p_1 \cdot p_2 + m_4^2$. To achieve the above expressions, we have chosen heavy quark momentum $\vec{p}_1$ along the $+z$ direction and the medium parton momentum $\vec{p}_2$ in the $xz$ plane ($\theta_2$ is the polar angle of $\vec{p}_2$). $(\theta_4, \phi_4)$ are the polar and azimuthal angle of $\vec{p}_4$. $\theta_{24}$ is the angle between $\vec{p}_2$ and $\vec{p}_4$. Due to the introduction of the thermal parton masses, the above expression is more complicated compared to the massless scenario in the earlier work~\cite{Cao:2016gvr}. Regarding the above two solutions, we choose the one which can return to the solution in Ref.~\cite{Cao:2016gvr} when we take the quasi-particles to be massless, i.e., the ``+" sign is taken for $A > 0$ and ``--” for $A < 0$. This is another modification by introducing the quasi-particle masses in addition to the matrix element and the density distribution discussed earlier.

In the LBT simulation, one first calculates the elastic scattering rate through each scattering channel $\Gamma_{\rm el}^{(i)}(\vec{p}, T)$ for a heavy quark with a given momentum $\vec{p}$ and a local temperature $T$ of the surrounding medium. The total scattering rate is the sum of different channels: $\Gamma_{\rm el} = \sum_{i} \Gamma_{\rm el}^{(i)}$.
Assuming the Poisson distribution for the number of scatterings in a given time step $\Delta t$, the elastic scattering probability can be given by:
\begin{align}
	P_{\rm el} = 1 - e^{-\Gamma_{\rm el} \Delta t}
\end{align}

\begin{figure*}[htb]
\includegraphics[width=0.375\linewidth]{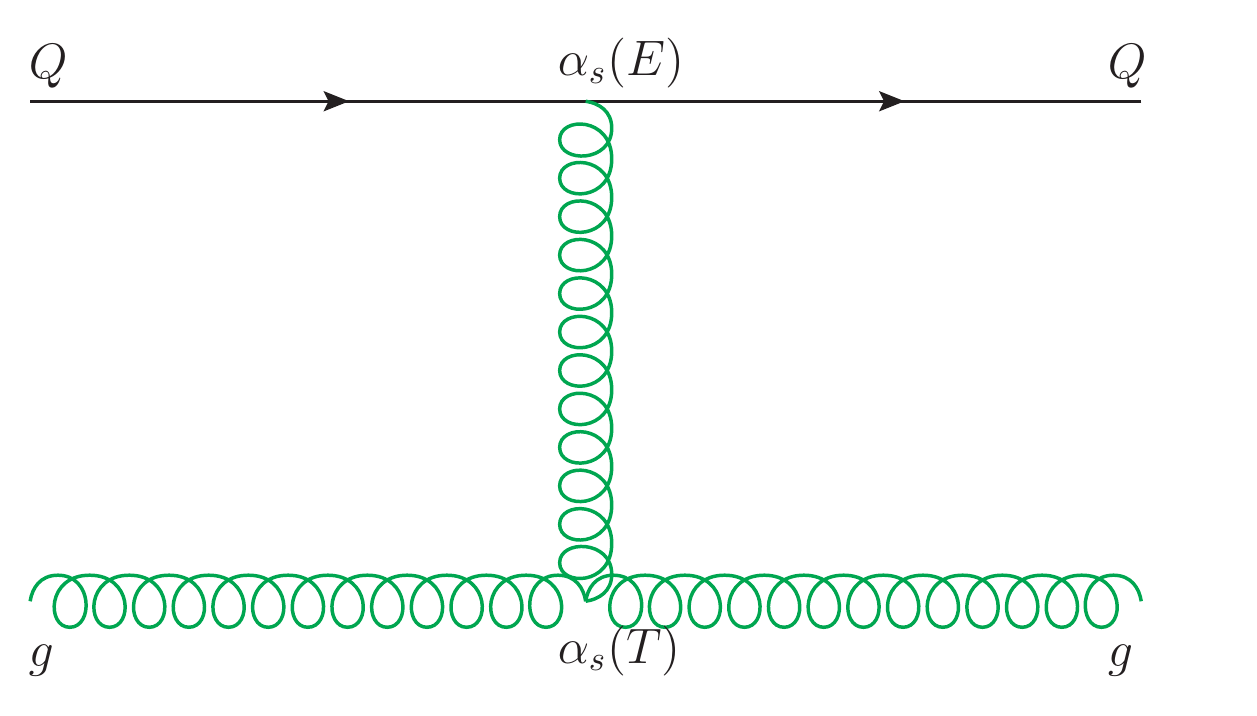}
\includegraphics[width=0.375\linewidth]{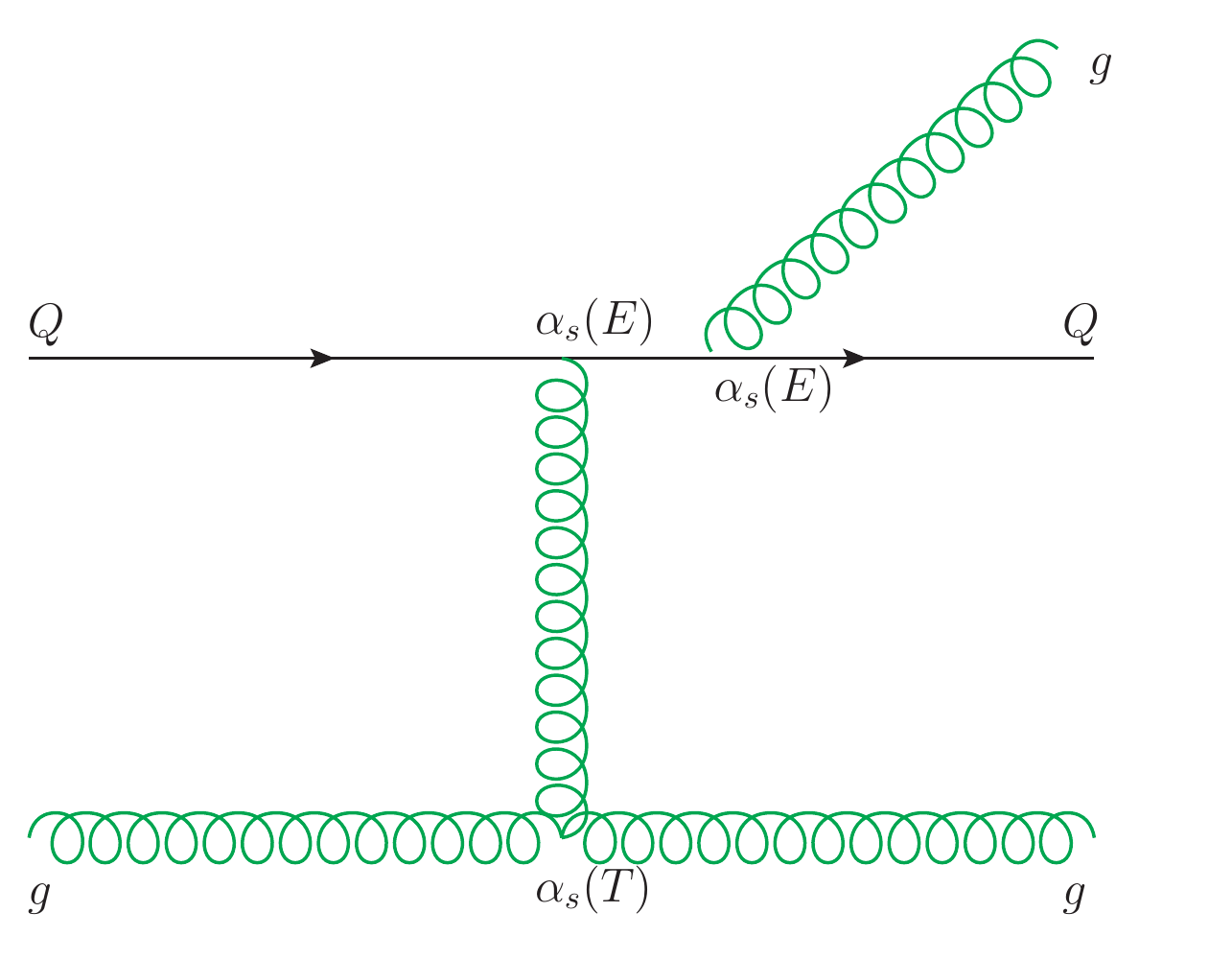}
\caption{Running couplings in elastic and inelastic scattering processes.}
\label{cg_el_inel}
\end{figure*}

For the inelastic process, or medium-induced gluon emission, experienced by parton ``1", we implement the medium-induced gluon spectra given by the higher-twist energy loss formalism ~\cite{Wang:2001ifa, Zhang:2003wk}:
\begin{align}
\frac{dN_g}{dx dk_\perp^2 dt}=\frac{2 \alpha_{\rm s} C_A P(x)k_\perp^4}{\pi (k_\perp^2+x^2M^2)^4} \hat{q} \sin^2 \bigg(\frac{t-t_i}{2\tau_f} \bigg),
\end{align}
where $x$ is the fractional energy of the emitted gluon taken from its parent parton, $k_\perp$ is the gluon transverse momentum with respect to the parent parton, $\alpha_\mathrm{s}$ is the strong coupling, $P(x)$ is the vacuum splitting function, $\hat{q}$ is the quark transport parameter characterizing the transverse momentum broadening rate due to elastic collisions between jet partons and medium constituents, $\tau_f$ is the formation time of the radiated gluon in the form of $\tau_f=2Ex (1-x)/(k_\perp^2+x^2 M^2)$ with $M$ being the mass of the parent parton, and $t_i$ denotes the initial time or the production time of the parent parton.

To simulate the gluon radiation process in a given time step $\Delta t$ in the LBT model, we first calculate the average gluon number from the above gluon spectrum as
\begin{align}
	\langle N_g \rangle(t, \Delta t) = \Gamma_{\rm inel} \Delta t = \Delta t \int dx dk_\perp^2 \frac{dN_g}{dx dk_\perp^2 dt},
\end{align}
in which a lower cut-off is imposed for the emitted gluon energy as $x_\mathrm{min}=\mu_\mathrm{D}/E$ in order to avoid possible divergence as $x\rightarrow 0$.
With the assumption of a Poisson distribution for the number of radiated gluons in a given time step $\Delta t$, the inelastic scattering probability can be written as
\begin{align}
P_\mathrm{inel}=1-e^{-\Gamma_{\rm inel}\Delta t}.
\end{align}

In the end, by combining elastic and inelastic processes, we obtain the following total scattering probability:
\begin{align}
	P_{\rm tot} = 1-e^{-(\Gamma_{\rm el}+\Gamma_{\rm inel})\Delta t} = P_{\rm el} + P_{\rm inel} - P_{\rm el} P_{\rm inel}.
\end{align}
The above probability can be splitted into two parts: pure elastic scattering with probability  $P_{\rm el} (1 - P_{\rm inel})$ and inelastic scattering with probability $P_{\rm inel}$.

One of the key quantity that determines the interaction strength between heavy quarks and the QGP medium is the strong coupling $\alpha_{\rm s}$. In fact, there are three different coupling parameters associated with the elastic and inelastic scatterings under discussion, which are in principle can be all different. As illustrated in Fig.~\ref{cg_el_inel}, we use $\alpha_\mathrm{s}=g^2(T)/(4\pi)$ for the vertex directly connecting to the thermal parton inside the medium, where $g(T)$ is the temperature dependent coupling parameter previously extracted from the lattice EoS. For the other two vertices that connect to the jet parton (heavy quark), we assume the following parametric form:
\begin{align}
\label{eq:paraAlphas}
\alpha_{\rm s}(E)&=\frac{12 \pi}{(11N_c - 2N_f) \log\left[\left(A  E/ T_\mathrm{c} + B\right)^2\right]},
\end{align}
where $E$ is the jet parton (heavy quark) energy, and the parameters $A$ and $B$ will be determined from the heavy quark observables, such as $R_\mathrm{AA}$ and $v_2$ of heavy mesons, in the next section.
In this work, the prior distributions for two parameters $A$ and $B$ are taken as uniform within the ranges summarized in Table \ref{tab:3}.

\begin{table}[htb]
\label{tab:3}
\centering
\vspace{-5pt}
\begin{tabular}{c|c|c}
 \hline
 Lattice Data & Parameters & Prior Range  \\
 \hline
  WB/HQ  & A & [0.032, 0.08]    \\
  WB/HQ  & B & [0.64, 15]  \\
  \hline
\end{tabular}
	\caption{The ranges of model parameters $(A,B)$ used in the prior distributions.}
\end{table}

\begin{figure*}[tbh]
\includegraphics[width=0.485\linewidth]{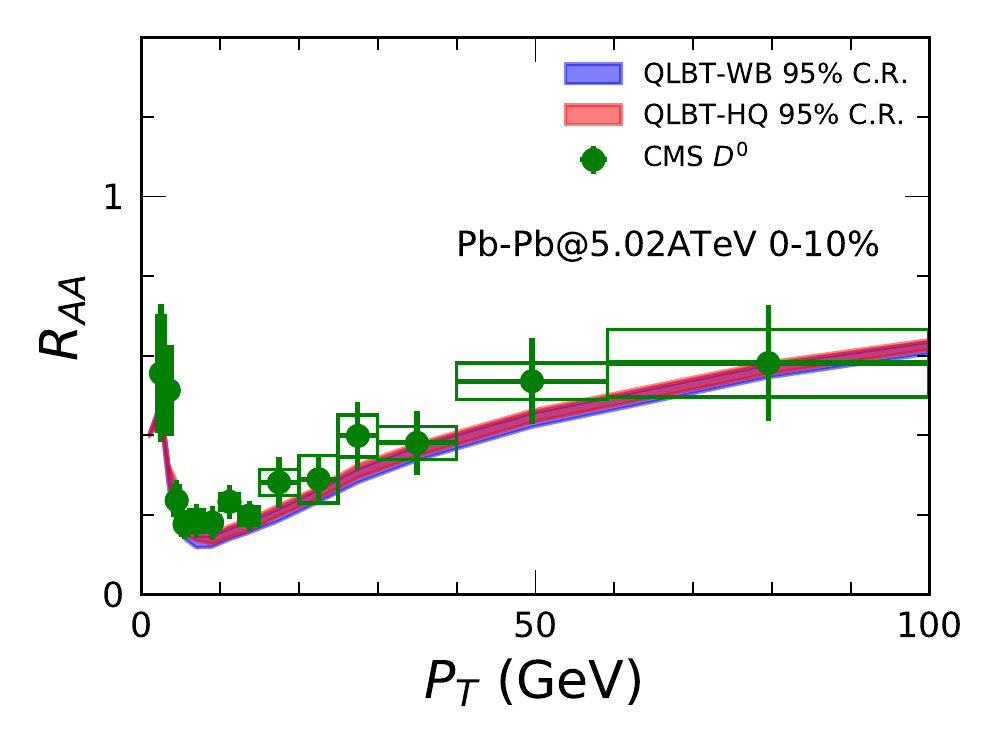}
\includegraphics[width=0.485\linewidth]{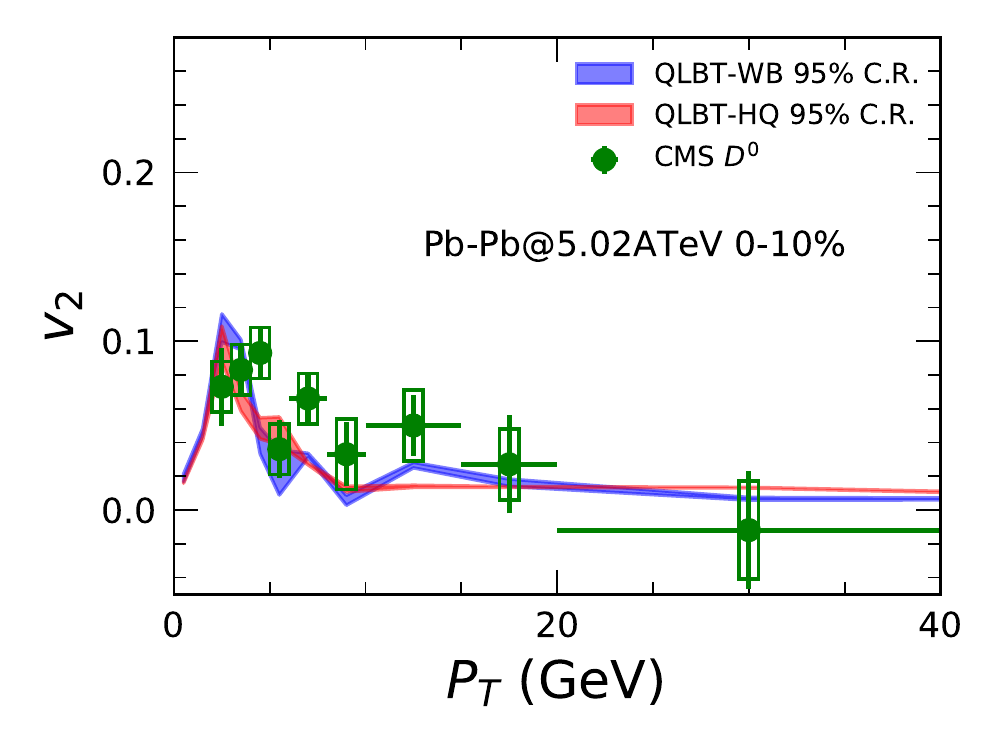}
\includegraphics[width=0.485\linewidth]{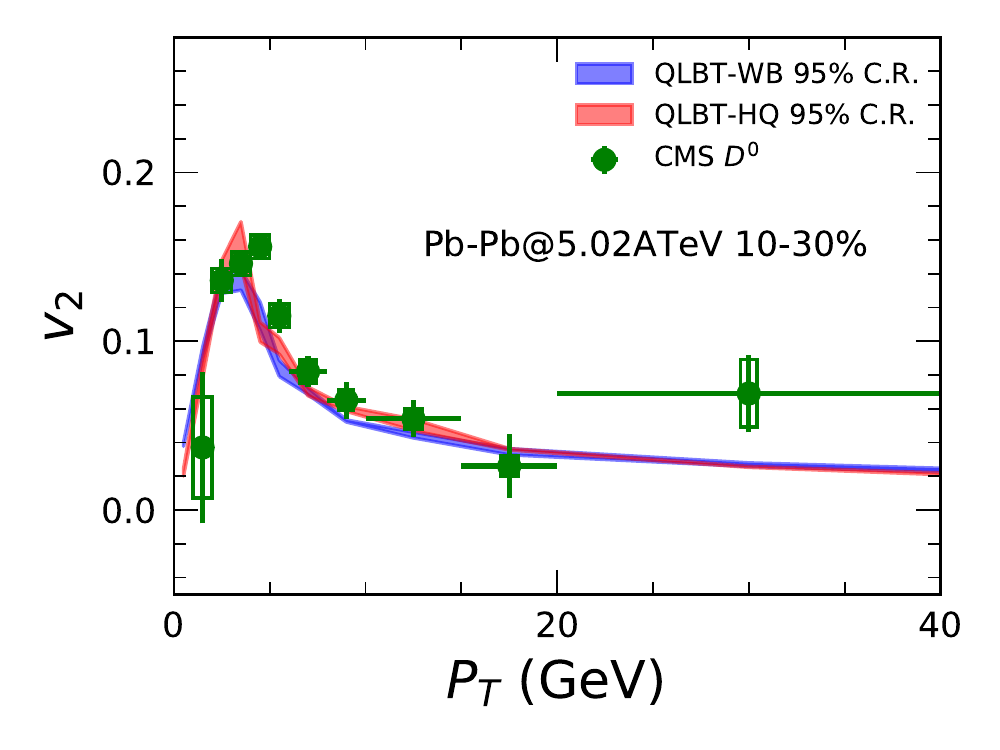}
\includegraphics[width=0.485\linewidth]{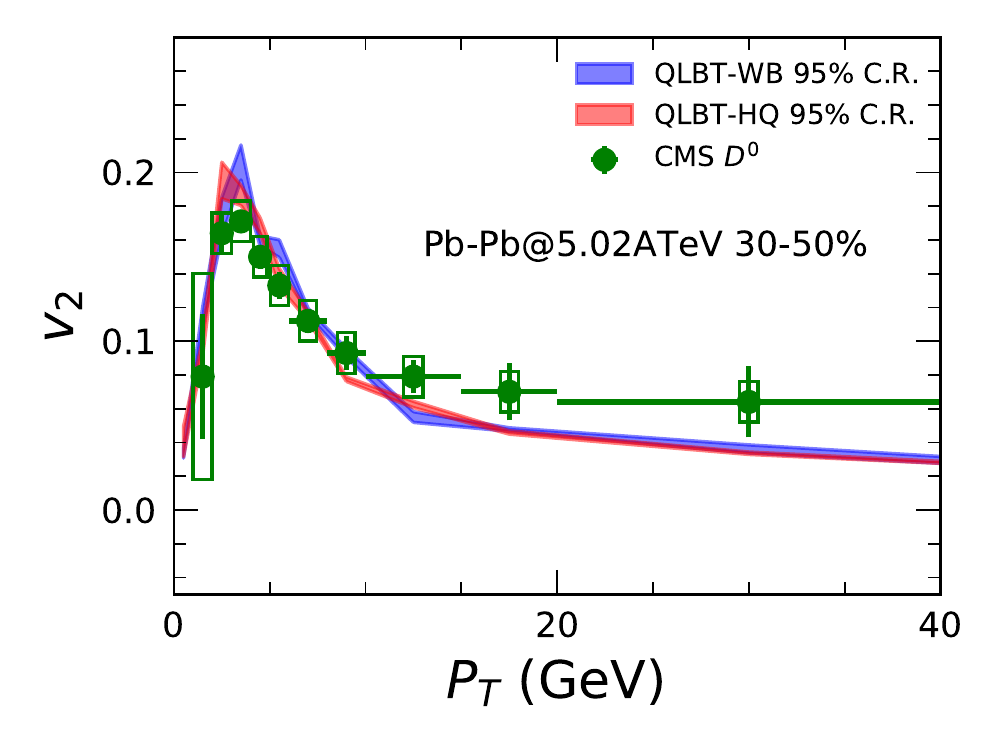}
\includegraphics[width=0.485\linewidth]{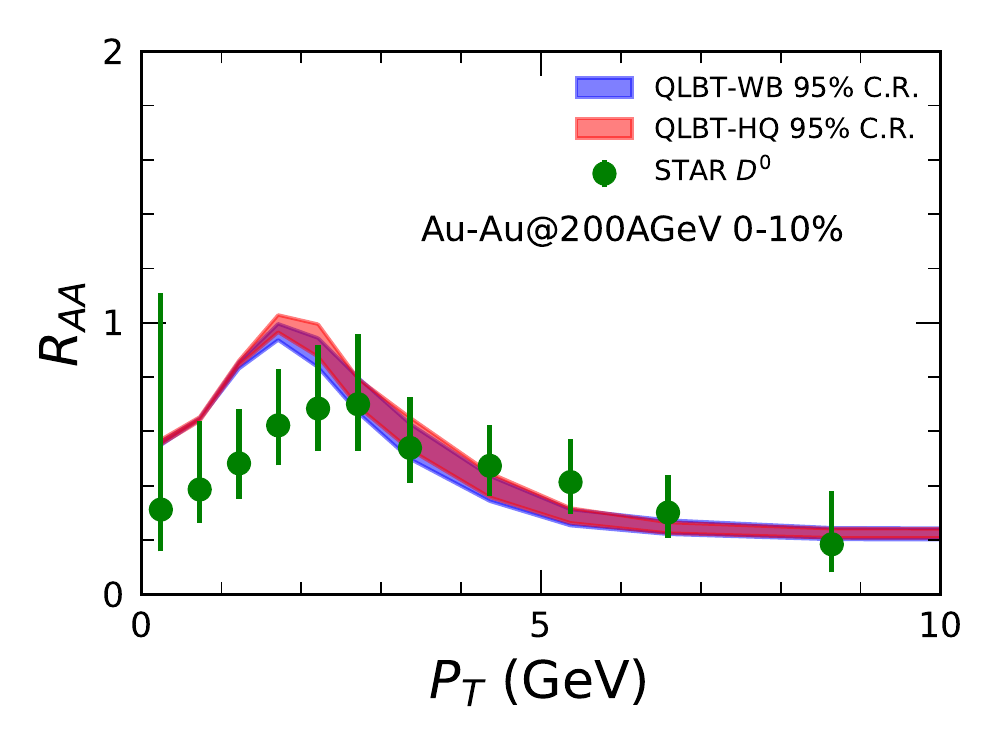}
\includegraphics[width=0.485\linewidth]{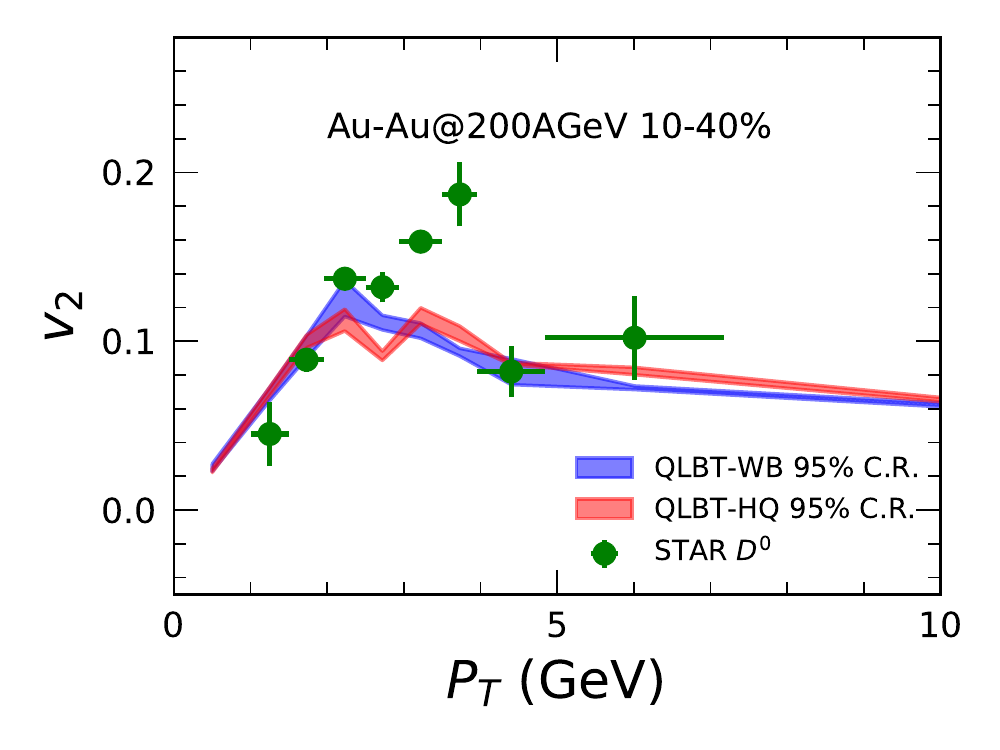}
	\caption{Using the Bayesian analysis, the nuclear modifiacation factors $R_\mathrm{AA}$ and the elliptic flow $v_2$ for $D$ mesons in Pb-Pb collisions at $\sqrt{s_{\rm NN}} = 5.02$~TeV at the LHC and in Au-Au collisions at $\sqrt{s_{\rm NN}}= 200$~GeV at RHIC are compared to the experimental data from CMS Collaboration~\cite{Sirunyan:2017xss, Sirunyan:2017plt} and STAR Collaboration~\cite{Adam:2018inb, Adamczyk:2017xur}.}
	\label{RAA-v2}
\end{figure*}

\begin{figure}[tbh]
\includegraphics[width=0.485\linewidth]{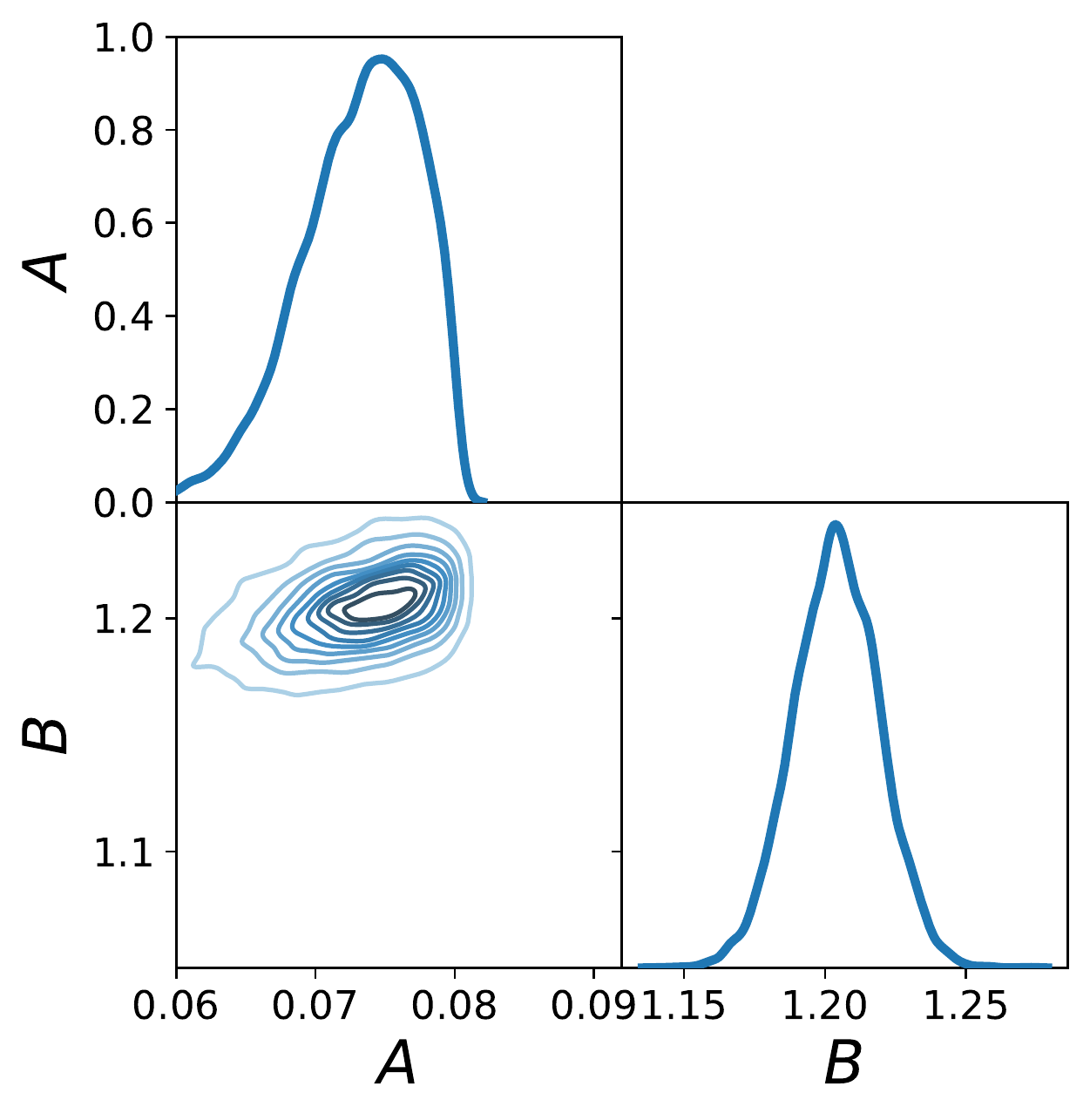}
\includegraphics[width=0.485\linewidth]{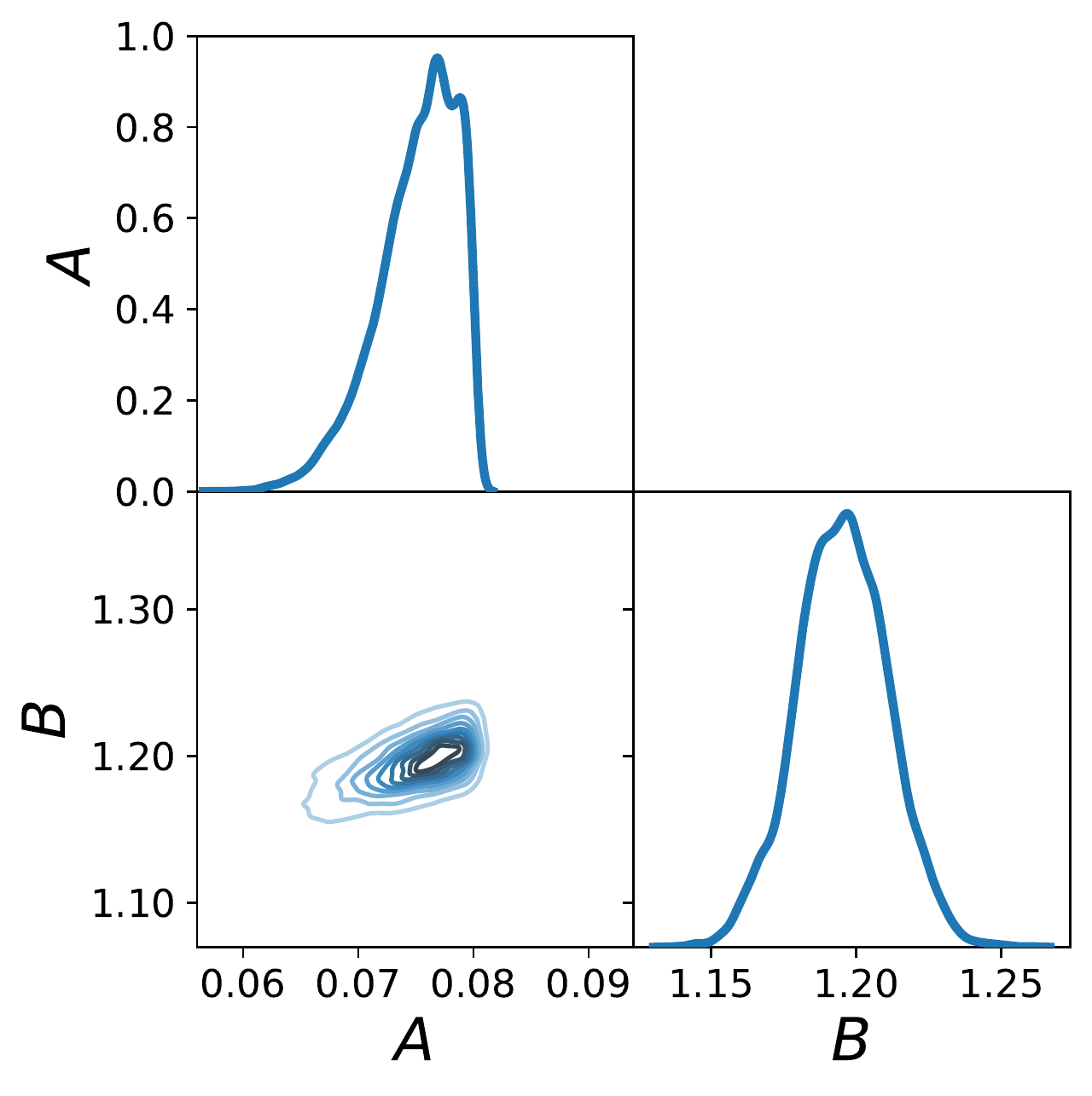}
	\caption{The posterior distribution of the parameters $A$ and $B$ in the strong coupling $\alpha_{\rm s}(E)$, using two different $g(T)$ obtained in Section II from fitting to WB (Left) and HQ (Right) lattice QCD data.}
	\label{AB}
\vspace{12pt}
\includegraphics[width=0.985\linewidth]{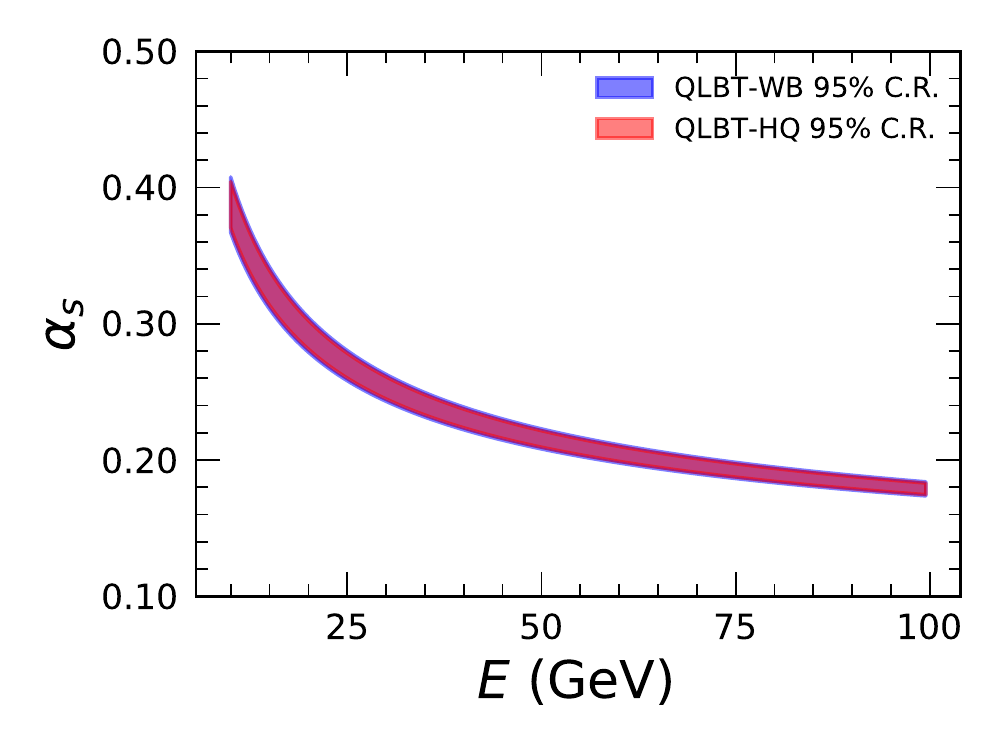}
	\caption{The energy-dependent running coupling $\alpha(E)$ obtained from the Bayesian analysis of $D$ meson $R_\mathrm{AA}$ and $v_2$, using two different $g(T)$ obtained in Section II from fitting to WB and HQ lattice QCD data.}
	\label{alphaE}
\end{figure}

\begin{figure*}[tbh]
\includegraphics[width=0.485\linewidth]{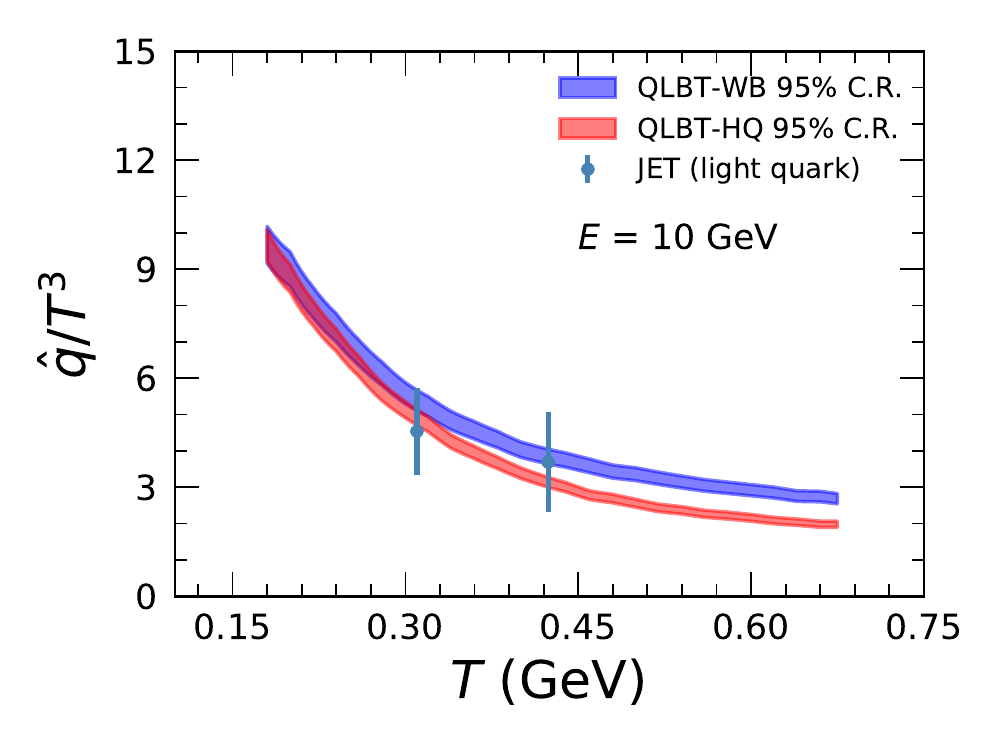}
\includegraphics[width=0.485\linewidth]{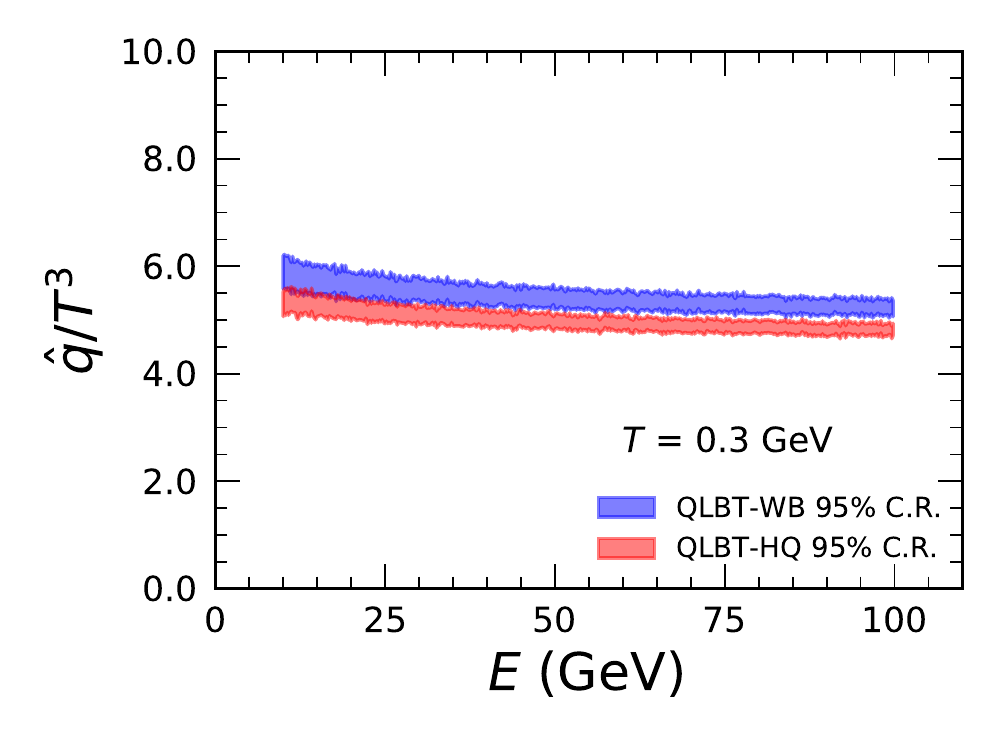}
\caption{ Temperature (Left) and energy (Right) dependent charm quark transport coefficient $\hat{q}/T^3$ obtained using two different $g(T)$ extracted from WB and HQ lattice QCD data. }
	\label{qhatT3}
\vspace{12pt}
\includegraphics[width=0.485\linewidth]{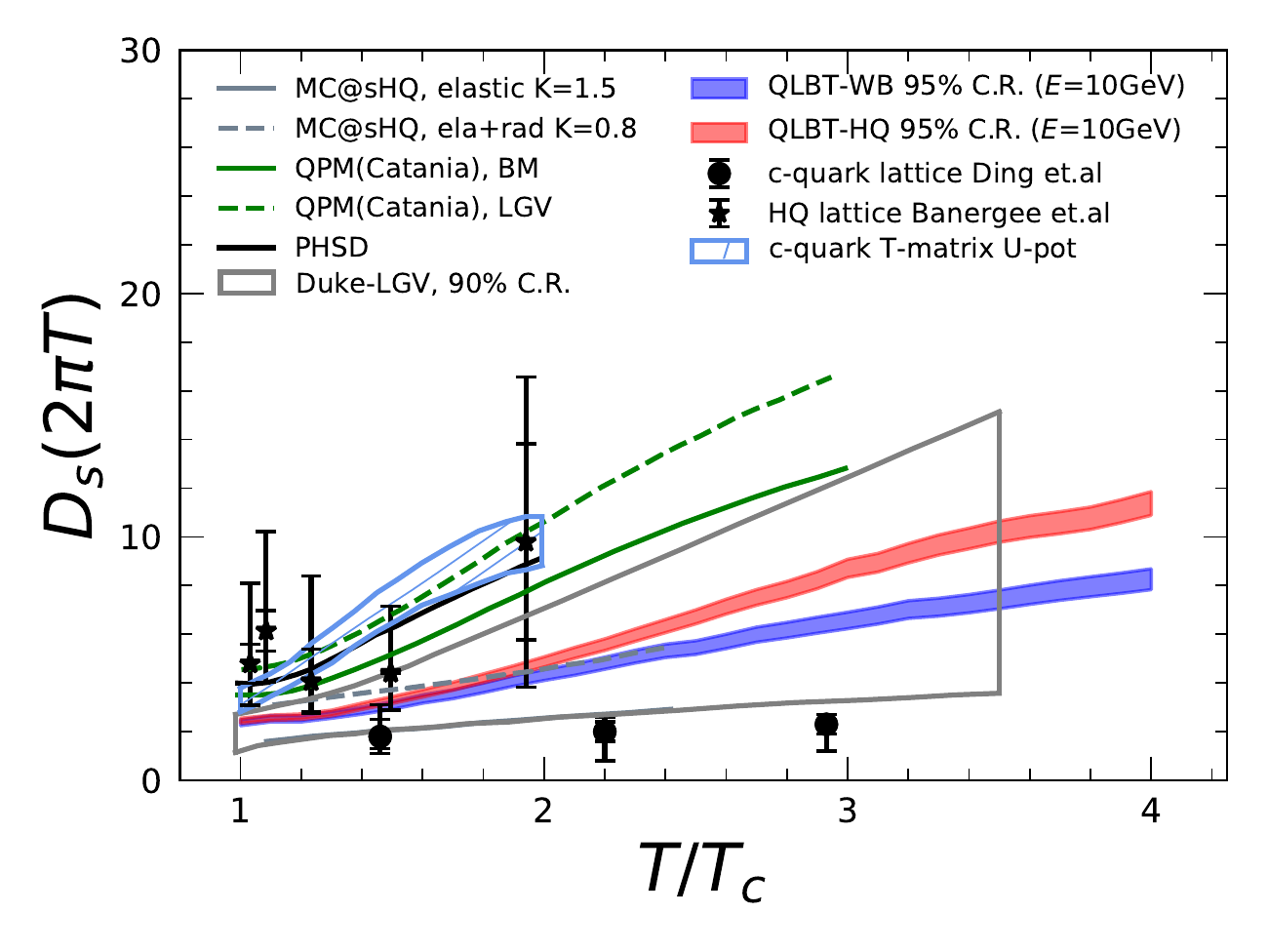}
\includegraphics[width=0.485\linewidth]{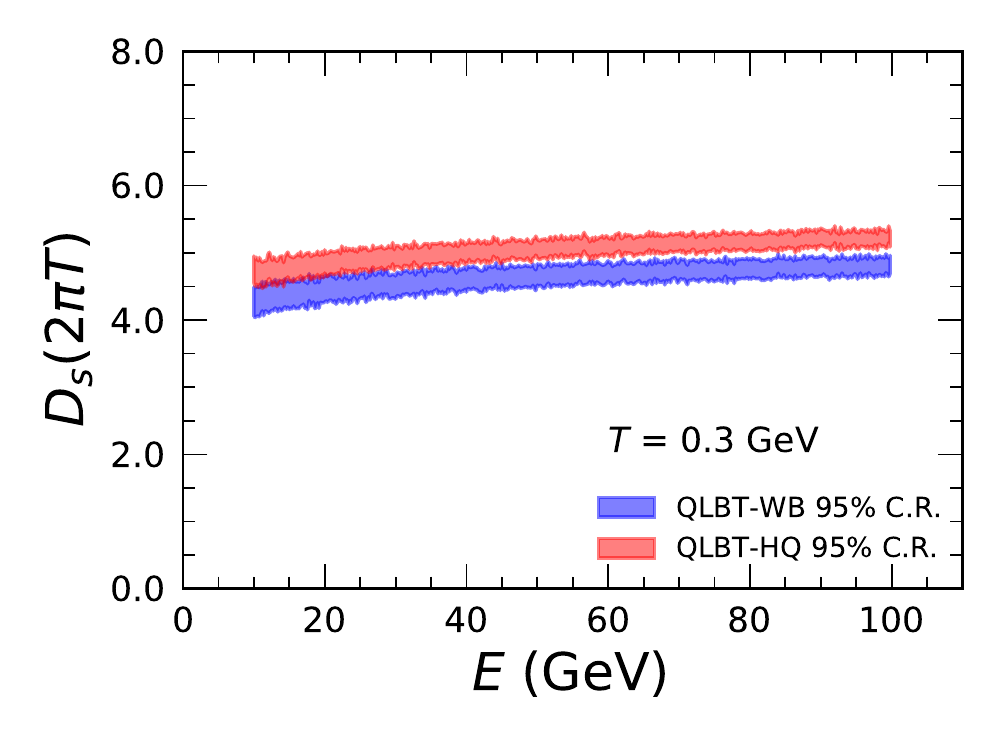}
\caption{Temperature (Left) and energy (Right) dependent charm quark diffusion coefficient $D_{\rm s}$ obtained using two different $g(T)$ extracted from WB and HQ lattice QCD data, compared to the results from different lattice QCD calculations~\cite{Ding:2012sp, Banerjee:2011ra} and phenomenological studies~\cite{Song:2015ykw,Scardina:2017ipo,Riek:2010fk,Gossiaux:2008jv,Xu:2017obm}.}
	\label{Ds2piT}
\end{figure*}

\section{NUMERICAL RESULTS for $R_{\rm AA}$ and $v_2$}
\label{sec:results}

Using the newly developed QLBT model as described above, one can calculate the nuclear modification factor $R_\mathrm{AA}$ and the elliptic flow $v_2$ for $D$ mesons at RHIC and the LHC. The spatial distributions of the heavy quark production vertices are calculated using the Monte-Carlo Glauber model, while their initial momentum distribution is taken from the LO perturbative QCD calculation that includes pair production ($q\bar{q}\rightarrow Q\bar{Q}$, $gg\rightarrow Q\bar{Q}$) and flavor excitation ($qQ\rightarrow qQ$, $gQ\rightarrow gQ$) processes. Meanwhile, the QGP medium is simulated using the (3+1)-dimensional CLVisc hydrodynamic model~\cite {Pang:2018zzo,Wu:2018cpc} whose initial energy density is obtained from the AMPT model~\cite{Lin:2004en}. In order to provide reasonable descriptions of the soft hadron spectra, the starting time of the hydrodynamic evolution is set as $\tau_0=0.6$~fm and the specific shear viscosity is set as $\eta/s=0.08$. 
Before the QGP phase ($\tau < \tau_0$), heavy quarks are assumed to stream freely. Possible interaction during this early stage \cite{Mrowczynski:2017kso, Carrington:2020sww} is neglected considering the short time duration of this stage compared to the ensuing QGP phase. The subsequent heavy quark interaction with the QGP is described using our QLBT model as described in the previous section. At the chemical freeze-out hypersurface ($T_\mathrm{c}$), we convert heavy quarks into heavy flavor hadrons using a hybrid coalescence-fragmentation hadronization model~\cite{Cao:2019iqs}, in which the coalescence probability between heavy quarks and thermal light quarks are determined by the wavefunction overlap between the free-quark state and hadronic bound state, and heavy quarks that do not hadronize through coalescence are fragmented into heavy flavor hadrons via Pythia~\cite{Sjostrand:2006za} simulation. Within this framework, we analyze the $R_\mathrm{AA}$ and $v_2$ of $D$ mesons in Pb-Pb collisions at $\sqrt{s_\mathrm{NN}} = 5.02$~TeV at the LHC and in Au-Au collisions at $\sqrt{s_\mathrm{NN}}= 200$~GeV at RHIC. The two model parameters $A$ and $B$ in Eq.~(\ref{eq:paraAlphas}) can then be obtained from calibrating the QLBT model to theses experimental data using the Bayesian statistical analysis method.
With the extracted parameters, one may further calculate heavy quark transport coefficient $\hat{q}$ and the spatial diffusion coefficient $D_\mathrm{s}$.

In Fig.~\ref{RAA-v2}, we first present the $D$ meson $R_\mathrm{AA}$ and $v_2$ at RHIC and the LHC after our model calibration. The error bands represent the the 95\% confidence interval. One can see that with the application of the lattice EoS, our QLBT model provides a reasonable description of the $D$ meson observables at both RHIC and the LHC. No significant difference can be observed between applying the WB EoS and the HQ EoS.

Shown in Fig.~\ref{AB} are the posterior distributions of the $(A, B)$ parameter space obtained from the above model calibration, in which the left and right panel corresponds to distribution extracted with WB and HQ EoS respectively. In the off-diagonal sub-figures, we also present the correlation between parameters $A$ and $B$.
Note that during the $\alpha_{\rm s}(E)$ calibration process, it is extremely computationally expensive to directly scan across the entire parameter space.
To speed up the computing efficiency~\cite{He:2018gks,Cao:2021keo}, we only apply the QLBT calculations on 50 sets of $(A, B)$ parameters that are sampled using the Latin-Hypercube algorithm, and then use the corresponding results to train the Gaussian process emulator. After being trained on the given design points, the emulator is able to interpolate between these design points and predict model results at an arbitrary point in the parameter space.
We then use the this Gaussian process emulator to scan across the parameter space, and update the sampled parameters using the MCMC algorithm together with the Bayesian statistics, in the same way that we applied to extract parameters of the EoS in Sec.~\ref{sec:QPM}.
In Fig.~\ref{AB}, one can observe that although two different lattice EoS are used in our analysis, they lead to similar values of $A$ and $B$ for heavy-quark-medium interaction coupling in the end (see their mean values and standard errors in Table \ref{tab:4}). This can be understood with the competing effects between the coupling strength $g(T)$ and the thermal parton mass $m(T)$ as shown in Fig.~\ref{gT-mT}. Although the WB EoS gives rise to a larger $g(T)$ than the HQ EoS does, the former yields a larger $m(T)$ than the latter as well. A larger $m(T)$ implies a lower parton density, thus a smaller scattering rate between heavy quarks and the QGP medium. These opposite effects from $g(T)$ and $m(T)$ cancel on the scattering rate, resulting in similar extracted values of the heavy-quark-medium interaction strength when comparing the QLBT model (with different EoS) to the experimental data.

\begin{table}[htb]
\label{tab:4}
\centering
\vspace{-5pt}
\begin{tabular}{c|c|c|c}
 \hline
 Lattice Data & Parameters & Mean Values & Standard Errors  \\
 \hline
  WB   & A & 0.073 & 0.004   \\
      & B & 1.204 & 0.015   \\ \hline
  HQ   & A & 0.075 & 0.0034  \\
  	  & B & 1.195 & 0.0157   \\
  \hline
\end{tabular}
	\caption{The mean and standard deviations of parameters $(A,B)$ in the coupling $\alpha_{\rm s}(E_Q)$ extracted from $R_{\rm AA}$ and $v_2$ using Wuppertal-Budapest and Hot QCD EOS.}
\end{table}

After obtaining the posterior distribution of the $(A, B)$ parameter space, we can straightforwardly calculate the strong coupling parameter $\alpha_\mathrm{s}$ using Eq.~(\ref{eq:paraAlphas}). In Fig.~\ref{alphaE}, we present the heavy quark energy dependent strong coupling $\alpha_{\rm s}(E)$. One can see that the coupling strength decreases as the heavy quark energy increases, which is crucial to explain the transverse momentum dependence of $D$ meson $R_\mathrm{AA}$. As discussed above, the coupling strength for heavy-quark-medium interaction has weak dependence on the choice of EoS (WB or HQ lattice QCD data) for our quasi-particle model.

In Fig.~\ref{qhatT3}, we further present the charm quark transport coefficient $\hat{q}$ obtained from our QLBT model, with its dependence on both the medium temperature (left) and the heavy quark energy (right). In the left plot, one can observe that the temperature-rescaled transport coefficient $\hat{q}/T^3$ decreases as the medium temperature $T$ increases. Our result is consistent with the ranges constrained by the earlier work from JET Collaboration~\cite{Burke:2013yra} for the RHIC and LHC energies separately. Another interesting feature one may observe is that $\hat{q}/T^3$ slightly decreases with the increase of the heavy quark energy $E$, as shown in the right plot. This is similar to the recent finding by JETSCAPE Collaboration~\cite{Cao:2021keo}; this is different from one's expectation from a direct perturbative calculation if a fixed coupling constant $\alpha_\mathrm{s}$ is utilized. Such energy dependence of $\hat{q}$ mainly originates from the strong energy dependence of $\alpha(E)$ as shown in Fig.~\ref{alphaE}. Although the heavy-quark-medium coupling strength in Fig.~\ref{alphaE} is not sensitive to the choice of EoS, after convoluting with the medium density and the coupling strength $g(T)$ inside the medium, sizable difference in $\hat{q}$ can be observed between using the WB and the HQ lattice QCD data, especially at high temperature.

Finally, we study the spatial diffusion coefficient $D_\mathrm{s}$. It can be related to the quark transport coefficient $\hat{q}$ via $D_\mathrm{s}(2\pi T) = 8\pi/(\hat{q}/T^3)$, where the relations $D_\mathrm{s}=T/(M\eta_\mathrm{D})$ and  $\eta_\mathrm{D}=\kappa/(2TE)$ are used, with $\eta_\mathrm{D}$ and $\kappa = \hat{q}/2$ known as the drag coefficient and momentum space diffusion coefficient respectively. Fig.~\ref{Ds2piT} shows the temperature and energy dependences of the diffusion coefficient $D_\mathrm{s}(2 \pi T)$. In the left plot, we also compare our result to the results from lattice QCD simulations \cite{Ding:2012sp, Banerjee:2011ra} and other phenomenological studies in literature \cite{Song:2015ykw,Scardina:2017ipo,Riek:2010fk,Gossiaux:2008jv,Xu:2017obm}.
Note that since $D$ meson $R_{\rm AA}$ and $v_2$ are not very sensitive to the values of heavy quark diffusion coefficient $D_{\rm s}$ at zero momentum, we present our $D_{\rm s}(T)$ for heavy quarks with $E=10$~GeV in Fig.~\ref{Ds2piT}.
One can see reasonable consistency between our results and other groups.
Since $D_\mathrm{s}(2\pi T)$ and $\hat{q}/T^3$ are inversely proportional to each other, we expect to see that $D_\mathrm{s}(2\pi T)$ increases as the medium temperature increases, and slightly increases as the heavy quark energy increases.

\section{SUMMARY}
\label{sec:summary}

In this work, we have performed a systematic study on heavy quark evolution and heavy meson production in relativistic heavy-ion collisions.
We have developed a new QLBT model based on the previous linear Boltzmann transport (LBT) model by treating the QGP as a collection of quasi-particles, whose temperature-dependent thermal masses $m(T)$ and interaction strength $g(T)$ have been obtained via calibrating the equation of state (EoS) -- pressure, entropy density and energy density -- of the quasi-particle system to the lattice QCD data using the Bayesian statistical analysis method.

By combining this new QLBT model with the CLVisc hydrodynamic model for the bulk evolution and the hybrid fragmentation-coalescence hadronization model that converts heavy quarks into heavy flavor hadrons, we have studied the medium modification of the heavy meson production in heavy-ion collisions at RHIC and the LHC. Through comparing the QLBT model results to the experimental data on the $D$ meson $R_\mathrm{AA}$ and $v_2$, we are able to extract the heavy-quark-QGP coupling strength $\alpha_\mathrm{s}$ as a function of the heavy quark energy using the Bayesian analysis method. Within this framework with 95\% confidence region, our $D$ meson observables are shown to be consistent with the experimental data. Meanwhile, our extracted heavy quark transport coefficients, including the quark transport coefficient $\hat{q}$ and the spatial diffusion coefficient $D_\mathrm{s}$, are shown to agree with earlier phenomenological studies in the literature as well as direct lattice QCD calculations. The sensitivity of the heavy flavor transport coefficients and observables to the EoS of the hot and dense nuclear matter has also been explored in this work. By comparing the EoS between the Wuppertal-Budapest (WB) and the Hot QCD (HQ) lattice QCD data, we have found that while the former yields a larger coupling strength $g(T)$ between quasi-particles, it provides larger quasi-particles masses $m(T)$ as well thus smaller thermal densities. These two effects can cancel each other and render similar amount of heavy quark energy loss at the end for two different choices of the EoS.
In the present study, we have shown that the final $D$ meson $R_\mathrm{AA}$ and $v_2$ as well as the extracted heavy-quark-QGP coupling strength $\alpha_\mathrm{s}(E)$ are insensitive to the choice of EoS between the WB and HQ data.
However, sizable differences can be seen in the extracted values of $\hat{q}$ and $D_\mathrm{s}$ transport coefficients.

\section*{Acknowledgments}

We thank Heng-Tong Ding for discussions. This work is supported in part by Natural Science Foundation of China under Grants No. 11775095, No. 11890710, No. 11890711 and No. 11935007, No. 11221504, No. 11861131009 and No. 11890714, by U.S. DOE under Grant No. DE-AC0205CH11231 within the JETSCAPE Collaboration, and by U.S. National Science Foundation under Grants No. ACI-1550228 and No. OAC-2004571 within XSCAPE Collaboration.
Some of the calculations were performed in the Nuclear Science Computing Center at Central China Normal University (NSC$^3$), Wuhan, Hubei, China.

\bibliographystyle{h-physrev5}
\bibliography{refs_GYQ}

\begin{thebibliography}{100}

\bibitem{Ollitrault:1992bk}
J.-Y. Ollitrault,
\newblock Phys. Rev. D {\bf 46}, 229 (1992).

\bibitem{Adler:2003kt}
PHENIX, S.~S. Adler {\em et~al.},
\newblock Phys. Rev. Lett. {\bf 91}, 182301 (2003), arXiv:nucl-ex/0305013.

\bibitem{Adams:2003am}
STAR, J.~Adams {\em et~al.},
\newblock Phys. Rev. Lett. {\bf 92}, 052302 (2004), arXiv:nucl-ex/0306007.

\bibitem{Aamodt:2010pa}
The ALICE Collaboration, K.~Aamodt {\em et~al.},
\newblock Phys.Rev.Lett. {\bf 105}, 252302 (2010), arXiv:1011.3914.

\bibitem{Gyulassy:1996br}
M.~Gyulassy, D.~H. Rischke, and B.~Zhang,
\newblock Nucl. Phys. A {\bf 613}, 397 (1997), arXiv:nucl-th/9609030.

\bibitem{Aguiar:2001ac}
C.~E. Aguiar, Y.~Hama, T.~Kodama, and T.~Osada,
\newblock Nucl. Phys. A {\bf 698}, 639 (2002), arXiv:hep-ph/0106266.

\bibitem{Broniowski:2007ft}
W.~Broniowski, P.~Bozek, and M.~Rybczynski,
\newblock Phys. Rev. {\bf C76}, 054905 (2007), arXiv:0706.4266.

\bibitem{Andrade:2008xh}
R.~P.~G. Andrade, F.~Grassi, Y.~Hama, T.~Kodama, and W.~L. Qian,
\newblock Phys. Rev. Lett. {\bf 101}, 112301 (2008), arXiv:0805.0018.

\bibitem{Hirano:2009ah}
T.~Hirano and Y.~Nara,
\newblock Phys. Rev. C {\bf 79}, 064904 (2009), arXiv:0904.4080.

\bibitem{Alver:2010gr}
B.~Alver and G.~Roland,
\newblock Phys. Rev. {\bf C81}, 054905 (2010), arXiv:1003.0194.

\bibitem{Petersen:2010cw}
H.~Petersen, G.-Y. Qin, S.~A. Bass, and B.~Muller,
\newblock Phys.Rev. {\bf C82}, 041901 (2010), arXiv:1008.0625.

\bibitem{Qin:2010pf}
G.-Y. Qin, H.~Petersen, S.~A. Bass, and B.~Muller,
\newblock Phys.Rev. {\bf C82}, 064903 (2010), arXiv:1009.1847.

\bibitem{Staig:2010pn}
P.~Staig and E.~Shuryak,
\newblock (2010), arXiv:1008.3139.

\bibitem{Teaney:2010vd}
D.~Teaney and L.~Yan,
\newblock Phys.Rev. {\bf C83}, 064904 (2011), arXiv:1010.1876.

\bibitem{Schenke:2010rr}
B.~Schenke, S.~Jeon, and C.~Gale,
\newblock Phys.Rev.Lett. {\bf 106}, 042301 (2011), arXiv:1009.3244.

\bibitem{Ma:2010dv}
G.-L. Ma and X.-N. Wang,
\newblock Phys.Rev.Lett. {\bf 106}, 162301 (2011), arXiv:1011.5249.

\bibitem{Qiu:2011iv}
Z.~Qiu and U.~W. Heinz,
\newblock Phys.Rev. {\bf C84}, 024911 (2011), arXiv:1104.0650.

\bibitem{Zhao:2019ehg}
W.~Zhao, C.~M. Ko, Y.-X. Liu, G.-Y. Qin, and H.~Song,
\newblock Phys. Rev. Lett. {\bf 125}, 072301 (2020), arXiv:1911.00826.

\bibitem{Heinz:2013th}
U.~Heinz and R.~Snellings,
\newblock Ann.Rev.Nucl.Part.Sci. {\bf 63}, 123 (2013), arXiv:1301.2826.

\bibitem{Gale:2013da}
C.~Gale, S.~Jeon, and B.~Schenke,
\newblock Int.J.Mod.Phys. {\bf A28}, 1340011 (2013), arXiv:1301.5893.

\bibitem{Huovinen:2013wma}
P.~Huovinen,
\newblock Int.J.Mod.Phys. {\bf E22}, 1330029 (2013), arXiv:1311.1849.

\bibitem{Bernhard:2019bmu}
J.~E. Bernhard, J.~S. Moreland, and S.~A. Bass,
\newblock Nature Phys. {\bf 15}, 1113 (2019).

\bibitem{JETSCAPE:2020shq}
JETSCAPE, D.~Everett {\em et~al.},
\newblock Phys. Rev. Lett. {\bf 126}, 242301 (2021), arXiv:2010.03928.

\bibitem{Khachatryan:2016odn}
CMS, V.~Khachatryan {\em et~al.},
\newblock JHEP {\bf 04}, 039 (2017), arXiv:1611.01664.

\bibitem{Acharya:2018qsh}
ALICE, S.~Acharya {\em et~al.},
\newblock JHEP {\bf 11}, 013 (2018), arXiv:1802.09145.

\bibitem{Aad:2015wga}
ATLAS, G.~Aad {\em et~al.},
\newblock JHEP {\bf 09}, 050 (2015), arXiv:1504.04337.

\bibitem{Burke:2013yra}
JET, K.~M. Burke {\em et~al.},
\newblock Phys. Rev. {\bf C90}, 014909 (2014), arXiv:1312.5003.

\bibitem{Buzzatti:2011vt}
A.~Buzzatti and M.~Gyulassy,
\newblock Phys. Rev. Lett. {\bf 108}, 022301 (2012), arXiv:1106.3061.

\bibitem{Chien:2015vja}
Y.-T. Chien, A.~Emerman, Z.-B. Kang, G.~Ovanesyan, and I.~Vitev,
\newblock Phys. Rev. {\bf D93}, 074030 (2016), arXiv:1509.02936.

\bibitem{Andres:2016iys}
C.~Andrés, N.~Armesto, M.~Luzum, C.~A. Salgado, and P.~Zurita,
\newblock Eur. Phys. J. {\bf C76}, 475 (2016), arXiv:1606.04837.

\bibitem{Cao:2017hhk}
S.~Cao, T.~Luo, G.-Y. Qin, and X.-N. Wang,
\newblock Phys. Lett. {\bf B777}, 255 (2018), arXiv:1703.00822.

\bibitem{Zigic:2018ovr}
D.~Zigic, I.~Salom, J.~Auvinen, M.~Djordjevic, and M.~Djordjevic,
\newblock Phys. Lett. {\bf B791}, 236 (2019), arXiv:1805.04786.

\bibitem{Wang:1991xy}
X.-N. Wang and M.~Gyulassy,
\newblock Phys.Rev.Lett. {\bf 68}, 1480 (1992).

\bibitem{Qin:2015srf}
G.-Y. Qin and X.-N. Wang,
\newblock Int. J. Mod. Phys. {\bf E24}, 1530014 (2015), arXiv:1511.00790.

\bibitem{Blaizot:2015lma}
J.-P. Blaizot and Y.~Mehtar-Tani,
\newblock Int. J. Mod. Phys. {\bf E24}, 1530012 (2015), arXiv:1503.05958.

\bibitem{Majumder:2010qh}
A.~Majumder and M.~Van~Leeuwen,
\newblock Prog.Part.Nucl.Phys. {\bf A66}, 41 (2011), arXiv:1002.2206.

\bibitem{Gyulassy:2003mc}
M.~Gyulassy, I.~Vitev, X.-N. Wang, and B.-W. Zhang,
\newblock (2003), arXiv:nucl-th/0302077.

\bibitem{Cao:2020wlm}
S.~Cao and X.-N. Wang,
\newblock Rept. Prog. Phys. {\bf 84}, 024301 (2021), arXiv:2002.04028.

\bibitem{Qin:2007rn}
G.-Y. Qin {\em et~al.},
\newblock Phys. Rev. Lett. {\bf 100}, 072301 (2008), arXiv:0710.0605.

\bibitem{Aad:2014bxa}
ATLAS, G.~Aad {\em et~al.},
\newblock Phys. Rev. Lett. {\bf 114}, 072302 (2015), arXiv:1411.2357.

\bibitem{Khachatryan:2016jfl}
CMS, V.~Khachatryan {\em et~al.},
\newblock Phys. Rev. {\bf C96}, 015202 (2017), arXiv:1609.05383.

\bibitem{Qin:2010mn}
G.-Y. Qin and B.~Muller,
\newblock Phys. Rev. Lett. {\bf 106}, 162302 (2011), arXiv:1012.5280,
\newblock [Erratum: Phys. Rev. Lett.108,189904(2012)].

\bibitem{Young:2011qx}
C.~Young, B.~Schenke, S.~Jeon, and C.~Gale,
\newblock Phys.Rev. {\bf C84}, 024907 (2011), arXiv:1103.5769.

\bibitem{Dai:2012am}
W.~Dai, I.~Vitev, and B.-W. Zhang,
\newblock Phys. Rev. Lett. {\bf 110}, 142001 (2013), arXiv:1207.5177.

\bibitem{Wang:2013cia}
X.-N. Wang and Y.~Zhu,
\newblock Phys. Rev. Lett. {\bf 111}, 062301 (2013), arXiv:1302.5874.

\bibitem{Blaizot:2013hx}
J.-P. Blaizot, E.~Iancu, and Y.~Mehtar-Tani,
\newblock Phys.Rev.Lett. {\bf 111}, 052001 (2013), arXiv:1301.6102.

\bibitem{Mehtar-Tani:2014yea}
Y.~Mehtar-Tani and K.~Tywoniuk,
\newblock Phys. Lett. {\bf B744}, 284 (2015), arXiv:1401.8293.

\bibitem{Cao:2017qpx}
S.~Cao and A.~Majumder,
\newblock (2017), arXiv:1712.10055.

\bibitem{Kang:2017frl}
Z.-B. Kang, F.~Ringer, and I.~Vitev,
\newblock Phys. Lett. {\bf B769}, 242 (2017), arXiv:1701.05839.

\bibitem{He:2018xjv}
Y.~He {\em et~al.},
\newblock (2018), arXiv:1809.02525.

\bibitem{Aad:2010bu}
Atlas Collaboration, G.~Aad {\em et~al.},
\newblock Phys.Rev.Lett. {\bf 105}, 252303 (2010), arXiv:1011.6182.

\bibitem{Chatrchyan:2012gt}
CMS, S.~Chatrchyan {\em et~al.},
\newblock Phys. Lett. {\bf B718}, 773 (2013), arXiv:1205.0206.

\bibitem{Qin:2009bk}
G.-Y. Qin, J.~Ruppert, C.~Gale, S.~Jeon, and G.~D. Moore,
\newblock Phys.Rev. {\bf C80}, 054909 (2009), arXiv:0906.3280.

\bibitem{Chen:2016vem}
L.~Chen, G.-Y. Qin, S.-Y. Wei, B.-W. Xiao, and H.-Z. Zhang,
\newblock Phys. Lett. {\bf B773}, 672 (2017), arXiv:1607.01932.

\bibitem{Chen:2016cof}
L.~Chen, G.-Y. Qin, S.-Y. Wei, B.-W. Xiao, and H.-Z. Zhang,
\newblock Phys. Lett. {\bf B782}, 773 (2018), arXiv:1612.04202.

\bibitem{Chen:2017zte}
W.~Chen, S.~Cao, T.~Luo, L.-G. Pang, and X.-N. Wang,
\newblock Phys. Lett. {\bf B777}, 86 (2018), arXiv:1704.03648.

\bibitem{Luo:2018pto}
T.~Luo, S.~Cao, Y.~He, and X.-N. Wang,
\newblock (2018), arXiv:1803.06785.

\bibitem{Zhang:2018urd}
S.-L. Zhang, T.~Luo, X.-N. Wang, and B.-W. Zhang,
\newblock Phys. Rev. {\bf C98}, 021901 (2018), arXiv:1804.11041.

\bibitem{Kang:2018wrs}
Z.-B. Kang, J.~Reiten, I.~Vitev, and B.~Yoon,
\newblock Phys. Rev. {\bf D99}, 034006 (2019), arXiv:1810.10007.

\bibitem{Chatrchyan:2013kwa}
CMS Collaboration, S.~Chatrchyan {\em et~al.},
\newblock Phys.Lett. {\bf B730}, 243 (2014), arXiv:1310.0878.

\bibitem{Aad:2014wha}
ATLAS, G.~Aad {\em et~al.},
\newblock Phys. Lett. {\bf B739}, 320 (2014), arXiv:1406.2979.

\bibitem{Chang:2016gjp}
N.-B. Chang and G.-Y. Qin,
\newblock Phys. Rev. {\bf C94}, 024902 (2016), arXiv:1603.01920.

\bibitem{Casalderrey-Solana:2016jvj}
J.~Casalderrey-Solana, D.~Gulhan, G.~Milhano, D.~Pablos, and K.~Rajagopal,
\newblock JHEP {\bf 03}, 135 (2017), arXiv:1609.05842.

\bibitem{Tachibana:2017syd}
Y.~Tachibana, N.-B. Chang, and G.-Y. Qin,
\newblock Phys. Rev. {\bf C95}, 044909 (2017), arXiv:1701.07951.

\bibitem{KunnawalkamElayavalli:2017hxo}
R.~Kunnawalkam~Elayavalli and K.~C. Zapp,
\newblock JHEP {\bf 07}, 141 (2017), arXiv:1707.01539.

\bibitem{Brewer:2017fqy}
J.~Brewer, K.~Rajagopal, A.~Sadofyev, and W.~Van Der~Schee,
\newblock JHEP {\bf 02}, 015 (2018), arXiv:1710.03237.

\bibitem{Chien:2016led}
Y.-T. Chien and I.~Vitev,
\newblock Phys. Rev. Lett. {\bf 119}, 112301 (2017), arXiv:1608.07283.

\bibitem{Milhano:2017nzm}
G.~Milhano, U.~A. Wiedemann, and K.~C. Zapp,
\newblock Phys. Lett. B {\bf 779}, 409 (2018), arXiv:1707.04142.

\bibitem{Chang:2019sae}
N.-B. Chang, Y.~Tachibana, and G.-Y. Qin,
\newblock Phys. Lett. {\bf B801}, 135181 (2020), arXiv:1906.09562.

\bibitem{Qin:2009uh}
G.~Y. Qin, A.~Majumder, H.~Song, and U.~Heinz,
\newblock Phys. Rev. Lett. {\bf 103}, 152303 (2009), arXiv:0903.2255.

\bibitem{Yang:2021iib}
Z.~Yang {\em et~al.},
\newblock (2021), arXiv:2101.05422.

\bibitem{Casalderrey-Solana:2020rsj}
J.~Casalderrey-Solana, J.~G. Milhano, D.~Pablos, K.~Rajagopal, and X.~Yao,
\newblock JHEP {\bf 05}, 230 (2021), arXiv:2010.01140.

\bibitem{Chen:2020tbl}
W.~Chen, S.~Cao, T.~Luo, L.-G. Pang, and X.-N. Wang,
\newblock Phys. Lett. B {\bf 810}, 135783 (2020), arXiv:2005.09678.

\bibitem{Yan:2017rku}
L.~Yan, S.~Jeon, and C.~Gale,
\newblock Phys. Rev. C {\bf 97}, 034914 (2018), arXiv:1707.09519.

\bibitem{Gao:2016ldo}
Z.~Gao, A.~Luo, G.-L. Ma, G.-Y. Qin, and H.-Z. Zhang,
\newblock Phys. Rev. C {\bf 97}, 044903 (2018), arXiv:1612.02548.

\bibitem{Cao:2021keo}
S.~Cao {\em et~al.},
\newblock (2021), arXiv:2102.11337.

\bibitem{Dong:2019byy}
X.~Dong, Y.-J. Lee, and R.~Rapp,
\newblock (2019), arXiv:1903.07709.

\bibitem{Rapp:2018qla}
A.~Beraudo {\em et~al.},
\newblock Nucl. Phys. {\bf A979}, 21 (2018), arXiv:1803.03824.

\bibitem{Cao:2018ews}
S.~Cao {\em et~al.},
\newblock (2018), arXiv:1809.07894.

\bibitem{Uphoff:2011ad}
J.~Uphoff, O.~Fochler, Z.~Xu, and C.~Greiner,
\newblock Phys. Rev. {\bf C84}, 024908 (2011), arXiv:1104.2295.

\bibitem{He:2011qa}
M.~He, R.~J. Fries, and R.~Rapp,
\newblock Phys. Rev. {\bf C86}, 014903 (2012), arXiv:1106.6006.

\bibitem{Young:2011ug}
C.~Young, B.~Schenke, S.~Jeon, and C.~Gale,
\newblock Phys. Rev. {\bf C86}, 034905 (2012), arXiv:1111.0647.

\bibitem{Alberico:2011zy}
W.~M. Alberico {\em et~al.},
\newblock Eur. Phys. J. {\bf C71}, 1666 (2011), arXiv:1101.6008.

\bibitem{Nahrgang:2013saa}
M.~Nahrgang, J.~Aichelin, P.~B. Gossiaux, and K.~Werner,
\newblock Phys. Rev. C {\bf 90}, 024907 (2014), arXiv:1305.3823.

\bibitem{Cao:2013ita}
S.~Cao, G.-Y. Qin, and S.~A. Bass,
\newblock Phys.Rev. {\bf C88}, 044907 (2013), arXiv:1308.0617.

\bibitem{Djordjevic:2013xoa}
M.~Djordjevic and M.~Djordjevic,
\newblock Phys. Lett. B {\bf 734}, 286 (2014), arXiv:1307.4098.

\bibitem{Cao:2015hia}
S.~Cao, G.-Y. Qin, and S.~A. Bass,
\newblock Phys. Rev. {\bf C92}, 024907 (2015), arXiv:1505.01413.

\bibitem{Das:2015ana}
S.~K. Das, F.~Scardina, S.~Plumari, and V.~Greco,
\newblock Phys. Lett. B {\bf 747}, 260 (2015), arXiv:1502.03757.

\bibitem{Song:2015ykw}
T.~Song, H.~Berrehrah, D.~Cabrera, W.~Cassing, and E.~Bratkovskaya,
\newblock Phys. Rev. C {\bf 93}, 034906 (2016), arXiv:1512.00891.

\bibitem{Cao:2016gvr}
S.~Cao, T.~Luo, G.-Y. Qin, and X.-N. Wang,
\newblock Phys. Rev. {\bf C94}, 014909 (2016), arXiv:1605.06447.

\bibitem{Kang:2016ofv}
Z.-B. Kang, F.~Ringer, and I.~Vitev,
\newblock JHEP {\bf 03}, 146 (2017), arXiv:1610.02043.

\bibitem{Prado:2016szr}
C.~A.~G. Prado {\em et~al.},
\newblock Phys. Rev. {\bf C96}, 064903 (2017), arXiv:1611.02965.

\bibitem{Cao:2017crw}
S.~Cao, A.~Majumder, G.-Y. Qin, and C.~Shen,
\newblock Phys. Lett. {\bf B793}, 433 (2019), arXiv:1711.09053.

\bibitem{Liu:2017qah}
S.~Y.~F. Liu and R.~Rapp,
\newblock Phys. Rev. {\bf C97}, 034918 (2018), arXiv:1711.03282.

\bibitem{Li:2018izm}
S.~Li, C.~Wang, X.~Yuan, and S.~Feng,
\newblock Phys. Rev. {\bf C98}, 014909 (2018), arXiv:1803.01508.

\bibitem{Ke:2018tsh}
W.~Ke, Y.~Xu, and S.~A. Bass,
\newblock Phys. Rev. {\bf C98}, 064901 (2018), arXiv:1806.08848.

\bibitem{Katz:2019fkc}
R.~Katz, C.~A.~G. Prado, J.~Noronha-Hostler, J.~Noronha, and A.~A.~P. Suaide,
\newblock (2019), arXiv:1906.10768.

\bibitem{Xing:2019xae}
W.-J. Xing, S.~Cao, G.-Y. Qin, and H.~Xing,
\newblock Phys. Lett. B {\bf 805}, 135424 (2020), arXiv:1906.00413.

\bibitem{Li:2020kax}
S.-Q. Li, W.-J. Xing, F.-L. Liu, S.~Cao, and G.-Y. Qin,
\newblock Chin. Phys. C {\bf 44}, 114101 (2020), arXiv:2005.03330.

\bibitem{He:2015pra}
Y.~He, T.~Luo, X.-N. Wang, and Y.~Zhu,
\newblock Phys. Rev. {\bf C91}, 054908 (2015), arXiv:1503.03313,
\newblock [Erratum: Phys. Rev.C97,no.1,019902(2018)].

\bibitem{Gorenstein:1995vm}
M.~I. Gorenstein and S.-N. Yang,
\newblock Phys. Rev. D {\bf 52}, 5206 (1995).

\bibitem{Levai:1997yx}
P.~Levai and U.~W. Heinz,
\newblock Phys. Rev. C {\bf 57}, 1879 (1998), arXiv:hep-ph/9710463.

\bibitem{Bozek:1998dj}
P.~Bozek, Y.~B. He, and J.~Hufner,
\newblock Phys. Rev. C {\bf 57}, 3263 (1998), arXiv:nucl-th/9802021.

\bibitem{Bluhm:2004xn}
M.~Bluhm, B.~Kampfer, and G.~Soff,
\newblock Phys. Lett. B {\bf 620}, 131 (2005), arXiv:hep-ph/0411106.

\bibitem{Plumari:2011mk}
S.~Plumari, W.~M. Alberico, V.~Greco, and C.~Ratti,
\newblock Phys. Rev. D {\bf 84}, 094004 (2011), arXiv:1103.5611.

\bibitem{Cassing:2008nn}
W.~Cassing,
\newblock Eur. Phys. J. ST {\bf 168}, 3 (2009), arXiv:0808.0715.

\bibitem{Cassing:2008sv}
W.~Cassing and E.~L. Bratkovskaya,
\newblock Phys. Rev. C {\bf 78}, 034919 (2008), arXiv:0808.0022.

\bibitem{Gossiaux:2009mk}
P.~B. Gossiaux, R.~Bierkandt, and J.~Aichelin,
\newblock Phys. Rev. C {\bf 79}, 044906 (2009), arXiv:0901.0946.

\bibitem{Cassing:2009vt}
W.~Cassing and E.~L. Bratkovskaya,
\newblock Nucl. Phys. A {\bf 831}, 215 (2009), arXiv:0907.5331.

\bibitem{Bratkovskaya:2011wp}
E.~L. Bratkovskaya, W.~Cassing, V.~P. Konchakovski, and O.~Linnyk,
\newblock Nucl. Phys. A {\bf 856}, 162 (2011), arXiv:1101.5793.

\bibitem{Berrehrah:2014kba}
H.~Berrehrah, P.-B. Gossiaux, J.~Aichelin, W.~Cassing, and E.~Bratkovskaya,
\newblock Phys. Rev. C {\bf 90}, 064906 (2014), arXiv:1405.3243.

\bibitem{Berrehrah:2015ywa}
H.~Berrehrah, E.~Bratkovskaya, W.~Cassing, P.~B. Gossiaux, and J.~Aichelin,
\newblock Phys. Rev. C {\bf 91}, 054902 (2015), arXiv:1502.01700.

\bibitem{Scardina:2017ipo}
F.~Scardina, S.~K. Das, V.~Minissale, S.~Plumari, and V.~Greco,
\newblock Phys. Rev. C {\bf 96}, 044905 (2017), arXiv:1707.05452.

\bibitem{Borsanyi:2013bia}
S.~Borsanyi {\em et~al.},
\newblock Phys. Lett. B {\bf 730}, 99 (2014), arXiv:1309.5258.

\bibitem{Bazavov:2014pvz}
HotQCD, A.~Bazavov {\em et~al.},
\newblock Phys. Rev. D {\bf 90}, 094503 (2014), arXiv:1407.6387.

\bibitem{He:2018gks}
Y.~He, L.-G. Pang, and X.-N. Wang,
\newblock Phys. Rev. Lett. {\bf 122}, 252302 (2019), arXiv:1808.05310.

\bibitem{pymc_bib}
J.~Salvatier, T.~Wiecki, and C.~Fonnesbeck,
\newblock Probabilistic programming in python using pymc, 2015,
  arXiv:1507.08050.

\bibitem{Auvinen:2009qm}
J.~Auvinen, K.~J. Eskola, and T.~Renk,
\newblock Phys. Rev. {\bf C82}, 024906 (2010), arXiv:0912.2265.

\bibitem{Combridge:1978kx}
B.~L. Combridge,
\newblock Nucl. Phys. B {\bf 151}, 429 (1979).

\bibitem{Wang:2001ifa}
X.-N. Wang and X.-f. Guo,
\newblock Nucl. Phys. {\bf A696}, 788 (2001), arXiv:hep-ph/0102230.

\bibitem{Zhang:2003wk}
B.-W. Zhang, E.~Wang, and X.-N. Wang,
\newblock Phys. Rev. Lett. {\bf 93}, 072301 (2004), arXiv:nucl-th/0309040.

\bibitem{Sirunyan:2017xss}
CMS, A.~M. Sirunyan {\em et~al.},
\newblock Phys. Lett. {\bf B782}, 474 (2018), arXiv:1708.04962.

\bibitem{Sirunyan:2017plt}
CMS, A.~M. Sirunyan {\em et~al.},
\newblock Phys. Rev. Lett. {\bf 120}, 202301 (2018), arXiv:1708.03497.

\bibitem{Adam:2018inb}
STAR, J.~Adam {\em et~al.},
\newblock Phys. Rev. C {\bf 99}, 034908 (2019), arXiv:1812.10224.

\bibitem{Adamczyk:2017xur}
STAR, L.~Adamczyk {\em et~al.},
\newblock Phys. Rev. Lett. {\bf 118}, 212301 (2017), arXiv:1701.06060.

\bibitem{Ding:2012sp}
H.~T. Ding {\em et~al.},
\newblock Phys. Rev. D {\bf 86}, 014509 (2012), arXiv:1204.4945.

\bibitem{Banerjee:2011ra}
D.~Banerjee, S.~Datta, R.~Gavai, and P.~Majumdar,
\newblock Phys. Rev. D {\bf 85}, 014510 (2012), arXiv:1109.5738.

\bibitem{Riek:2010fk}
F.~Riek and R.~Rapp,
\newblock Phys. Rev. C {\bf 82}, 035201 (2010), arXiv:1005.0769.

\bibitem{Gossiaux:2008jv}
P.~B. Gossiaux and J.~Aichelin,
\newblock Phys. Rev. C {\bf 78}, 014904 (2008), arXiv:0802.2525.

\bibitem{Xu:2017obm}
Y.~Xu, J.~E. Bernhard, S.~A. Bass, M.~Nahrgang, and S.~Cao,
\newblock Phys. Rev. C {\bf 97}, 014907 (2018), arXiv:1710.00807.

\bibitem{Pang:2018zzo}
L.-G. Pang, H.~Petersen, and X.-N. Wang,
\newblock Phys. Rev. {\bf C97}, 064918 (2018), arXiv:1802.04449.

\bibitem{Wu:2018cpc}
X.-Y. Wu, L.-G. Pang, G.-Y. Qin, and X.-N. Wang,
\newblock Phys. Rev. C {\bf 98}, 024913 (2018), arXiv:1805.03762.

\bibitem{Lin:2004en}
Z.-W. Lin, C.~M. Ko, B.-A. Li, B.~Zhang, and S.~Pal,
\newblock Phys.Rev. {\bf C72}, 064901 (2005), arXiv:nucl-th/0411110.

\bibitem{Mrowczynski:2017kso}
S.~Mrowczynski,
\newblock Eur. Phys. J. A {\bf 54}, 43 (2018), arXiv:1706.03127.

\bibitem{Carrington:2020sww}
M.~E. Carrington, A.~Czajka, and S.~Mrowczynski,
\newblock Nucl. Phys. A {\bf 1001}, 121914 (2020), arXiv:2001.05074.

\bibitem{Cao:2019iqs}
S.~Cao {\em et~al.},
\newblock Phys. Lett. B {\bf 807}, 135561 (2020), arXiv:1911.00456.

\bibitem{Sjostrand:2006za}
T.~Sjostrand, S.~Mrenna, and P.~Z. Skands,
\newblock JHEP {\bf 05}, 026 (2006), arXiv:hep-ph/0603175.

\end{thebibliography}
\end{document}